\documentclass[10pt,a4paper]{article}
\usepackage{f1000_styles}
\usepackage{multirow}
\usepackage{upgreek}

\usepackage[numbers]{natbib}

\begin{document}
\pagestyle{fancy}

\hyphenation{energy projection}
\title{Micromegas with GEM preamplification for enhanced energy threshold in low-background gaseous time projection chambers}

\author{Juan Castel, Susana Cebrián, Theopisti Dafni, David Díez-Ibáñez, Javier Galán, Juan Antonio García, Álvaro Ezquerro, Igor G. Irastorza, Gloria Luzón, Cristina Margalejo, Héctor Mirallas, Luis Obis, Alfonso Ortiz de Solórzano, Óscar Pérez, Jorge Porrón,  María J. Puyuelo}
\affil{Centro de Astropartículas y Física de Altas Energías, Universidad de Zaragoza, 50009 Zaragoza, Spain}

\maketitle
\thispagestyle{fancy}

\begin{abstract}

\textbf{Background:} We develop the concept of a Micromegas readout plane with an additional GEM preamplification stage placed a few millimetres above it to increase the maximum effective gain of the combined readout.\\ 
\textbf{Methods:} We implement it and test it in realistic conditions for its application to low-background dark matter searches like the TREX-DM experiment. For this, we use a Micromegas of Microbulk type, built with radiopure materials.\\
\textbf{Results:} We report on GEM effective extra gain factors of about 90, 50 and 20 in 1, 4 and 10~bar of Ar-1\%iC$_{4}$H$_{10}$. These results are obtained in a small test chamber allowing for systematic scanning of voltages and pressures. In addition, a TREX-DM full-scale set-up has also been built and tested, featuring a replica of the fully-patterned TREX-DM Microbulk readout.\\
\textbf{Conclusions:} The results here obtained show promise to lower the threshold of the experiment down to 50~eV$_{ee}$, corresponding to substantially enhanced sensitivity to low-mass WIMPs.

\end{abstract}

\section*{\color{OREblue}Keywords}

Dark Matter, WIMPs, Time Projection Chamber, Micromegas, Underground Science, Low Background Techniques, Radiopurity

\clearpage
\pagestyle{fancy}

\section{Introduction}
\label{sec:introduction}

Gaseous time projection chambers (TPCs) are versatile detectors that can measure the energy and track of ionizing particles in three dimensions. They have been widely used in various fields of physics, such as high-energy physics, nuclear physics, astroparticle physics and medical imaging~\cite{Hilke:2010zz}. One of the main advantages of TPCs is their ability to discriminate different types of particles based on their energy loss and track shape, which is crucial for reducing backgrounds in rare event searches. Modern incarnations of TPCs include micro-pattern gaseous detectors (MPGDs) as sensing planes. Example of these are the MICRO-MEsh GAseous Structure (Micromegas)~\cite{Giomataris:1995fq} and Gas Electron Amplifiers (GEM)~\cite{Sauli:1997,Sauli:2014}. In recent years, TPCs with Micromegas readouts have also found significant applications beyond traditional physics research, including muography for geophysics, civil engineering, and archaeological studies~\cite{Gomez:2019,Roche:2021}, demonstrating their utility across diverse scientific disciplines.

Micromegas is a type of MPGD that consists of a thin metallic mesh placed O(50)~$\upmu$m above a segmented anode. The gas volume between the cathode and the mesh acts as a drift region, while the gas volume between the mesh and the anode acts as an amplification region. The electric field in the amplification region is much higher than in the drift region, creating an avalanche of electrons that induces a signal on the anode~\cite{Giomataris:1995fq}.

In Micromegas of the ``Microbulk'' type, the whole amplification structure is produced by chemical processing of a double-sided copper-clad polyamide laminate, onto which the mesh and the anode pattern are etched ~\cite{Andriamonje:2010}. This type of detector is particularly well suited for low-background experiments thanks to its low intrinsic radioactivity~\cite{Cebrian:2010}. Indeed, it has been intensely developed in dedicated R\&D projects~\cite{T-REX_1:2016,T-REX_2:2016} and is now being used in solar axions~\cite{MicrobulkCAST_2010,IAXO_FerrerRibas:2023}, neutrinoless double beta decay~\cite{PandaX-III:2017} and direct dark matter searches~\cite{TREX-DM:2016,TREX-DM:2020,MicromegasPerspectives:2024}.

Although achieving higher signal-to-noise ratio is always a desirable property for every application, low detection threshold is especially relevant in dark matter experiments aiming at the detection of the low-energy nuclear recoils produced by the collisions of galactic WIMPs with target nuclei. In these experiments, the spectral distribution of the signal concentrates at low recoil energy, exponentially decaying for higher values~\cite{Lewin:1996}. For experiments specifically targeting low-mass WIMPs, lowering the detector threshold is an important line of detector development, that automatically translates to better sensitivity, while allowing access to lower WIMP masses. 

The intrinsic signal amplification happening in the Micromegas gap of the TPC, effectively decoupling the detector threshold from the total size of the TPC, is one of the appealing features that motivates the application of this technology in the TREX-DM experiment. TREX-DM~\cite{TREX-DM:2016} is a high-pressure, low-background, Micromegas-based TPC looking for low-mass WIMPs in the Canfranc Underground Laboratory under the Spanish Pyrenees. The TREX-DM TPC has been designed to have an active volume of 20~L, which translates into 0.32~kg of Argon mass at 10~bar (or, alternatively, 0.16~kg of Neon). It is composed of a cylindrical vessel made of radiopure copper, with a diameter of 0.5~m, a length of 0.5~m and a wall thickness of 6~cm. These dimensions are set by the requirements that the vessel holds up to 10~bar(a) of pressure, while at the same time constitutes the innermost part of the shielding. The vessel is divided into two active volumes by a central Mylar cathode, which is connected to high voltage by a tailor-made feedthrough. At each side, there is a 16-cm-long field cage defined by a series of copper strips imprinted on a Kapton substrate supported by four PTFE walls. At the two ends of the active volumes, two 25x25~cm$^2$ squared Microbulk Micromegas readouts are placed as sensing anodes, each of them patterned with $\sim$ 1~mm pixels, interlinked with 512 strips in an $x-y$ layout.

Especially optimized Micromegas test set-ups have shown that very high gains, of even >10$^6$, are achievable~\cite{Derre:2000}. However, the constraints imposed by the environment of a real experiment (the need for stable operation over long periods, robustness of operation, total absence of destructive discharges, large area, large readout segmentation, a controlled level of electronic noise, a given gas composition and pressure determined by physics, etc.) means that an energy threshold only somewhat lower than 1~keV is realistic for a Microbulk readout in an experiment like TREX-DM. Indeed, the target threshold of the experiment in its baseline configuration is 0.4~keV. The huge physics potential of lowering this parameter (potentially down to the single-electron level, $\sim$ 20~eV) in terms of improved sensitivity (see \autoref{sec:discussion}) has prompted the investigation to increase the operational gain of the readout, by a preamplification stage that multiplies the primary electron cloud before entering the Microbulk gap, effectively contributing with an additional multiplication factor to the final readout gain. This preamplification stage consists of a Gas Electron Multiplication (GEM) foil, which is made of a copper-clad (on both sides), 50-$\upmu$m-thick Kapton foil, perforated by a high density, regular matrix of holes. The primary electrons go through the holes and get multiplied by a factor depending on the voltage applied between the electrodes. The raw materials of GEM foils are identical to the ones of Microbulk planes, which makes this option promising for low-background searches.

In this article, we report on a study of a hybrid GEM plus Microbulk readout. The combination of a GEM and a Micromegas (GEM + MM henceforth) has been tested in the past~\cite{GEM+MM:2002} and successfully implemented in real experiments such as COMPASS~\cite{Neyret:2009,Neyret:2024} using bulk technology, but this is the first time this is done with a Microbulk Micromegas, at high pressures, and in the context of low-background constraints.

In order to perform the study, a test set-up has been built and operated, as described in \autoref{sec:description_test_set-up}. The results of the characterization of the combined readout are presented in \autoref{sec:results_test_set-up}. Later on, another set-up with a full-scale GEM foil installed on top of an exact replica of the TREX-DM Microbulks, simulating the real installation to be done in the TREX-DM experiment, is prepared to demonstrate the feasibility of this solution to enhance the threshold of the experiment. The description and results with this larger-scale set-up are presented in \autoref{sec:description_full-scale_set-up} and \autoref{sec:results_full-scale_set-up}. We briefly discuss in \autoref{sec:discussion} the sensitivity projections that the improvement in threshold of this work could potentially bring to an enhanced TREX-DM experiment. We finish with our conclusions in \autoref{sec:conclusions}.

\section{Methods}
\label{sec:methods}

This section is devoted to the description of the experimental set-ups used to obtain the results presented in this article.

\subsection{Description of the test set-up}
\label{sec:description_test_set-up}

The Microbulk Micromegas detector with the GEM preamplification stage is placed inside a small (2.4~L) stainless-steel chamber certified to withstand 12~bar. The vacuum level achieved in this chamber after $ \sim 1$~h with a Pfeiffer Vacuum HiCube 80 Classic Turbo Pump is $ \sim 10^{-5} $~mbar. The gas used in these studies is Ar-1\%iC$_{4}$H$_{10}$, though it is intended to extend them to other Ar- and Ne-based mixtures of interest to TREX-DM (such as Ar-10\%iC$_{4}$H$_{10}$).

The Microbulk Micromegas detector lies on top of a metallic support plate, separated from it by a PTFE piece. The Micromegas used has a non-segmented and disc-shaped anode with a small 2-cm-diameter circular active area (cathode). The gap between mesh and anode is 50~$ \upmu $m. Several Micromegas with varying hole diameters (50-60~$ \upmu $m) and hole pitch (100-110~$ \upmu $m) are used in these tests. A GEM stage of roughly the same active area is mounted on top of the mesh, at a distance $ L_{\mathrm{transfer}} = $ 10~mm. The GEM has a thickness of 60~$ \upmu $m (50~$ \upmu $m the Kapton, 5~$ \upmu $m each copper layer), hole pitch of 140~$ \upmu $m, diameter of holes in copper of 70~$ \upmu $m and diameter of holes in Kapton of 60~$ \upmu $m. Finally, a cathode (a stainless-steel grid) is placed above the GEM, at a distance $ L_{\mathrm{drift}} = $ 13~mm. The cathode has a $ ^{55} $Fe source attached (K-alpha X-ray at 5.9~keV) facing the ionization volume.

As for the voltages, the anode is kept grounded, the mesh at $ V_{\mathrm{mesh}} $, the bottom and top layers of the GEM at $ V_{\mathrm{bottom}} $ and $ V_{\mathrm{top}} $, respectively (we define the GEM preamplification voltage as $ V_{\mathrm{GEM}} = V_{\mathrm{top}}-V_{\mathrm{bottom}} $), and the cathode at $ V_{\mathrm{cath}} $. The metallic support plate is kept at $ V_{\mathrm{plate}}=V_{\mathrm{mesh}} $ to avoid potential distortions in the transfer field $ E_{\mathrm{transfer}} $. Two CAEN HV power supply modules (a 4-channel N1471H and a 2-channel N471A) are used to provide these voltages.

\begin{figure}
	\centering
	\includegraphics[width=0.5\textwidth]{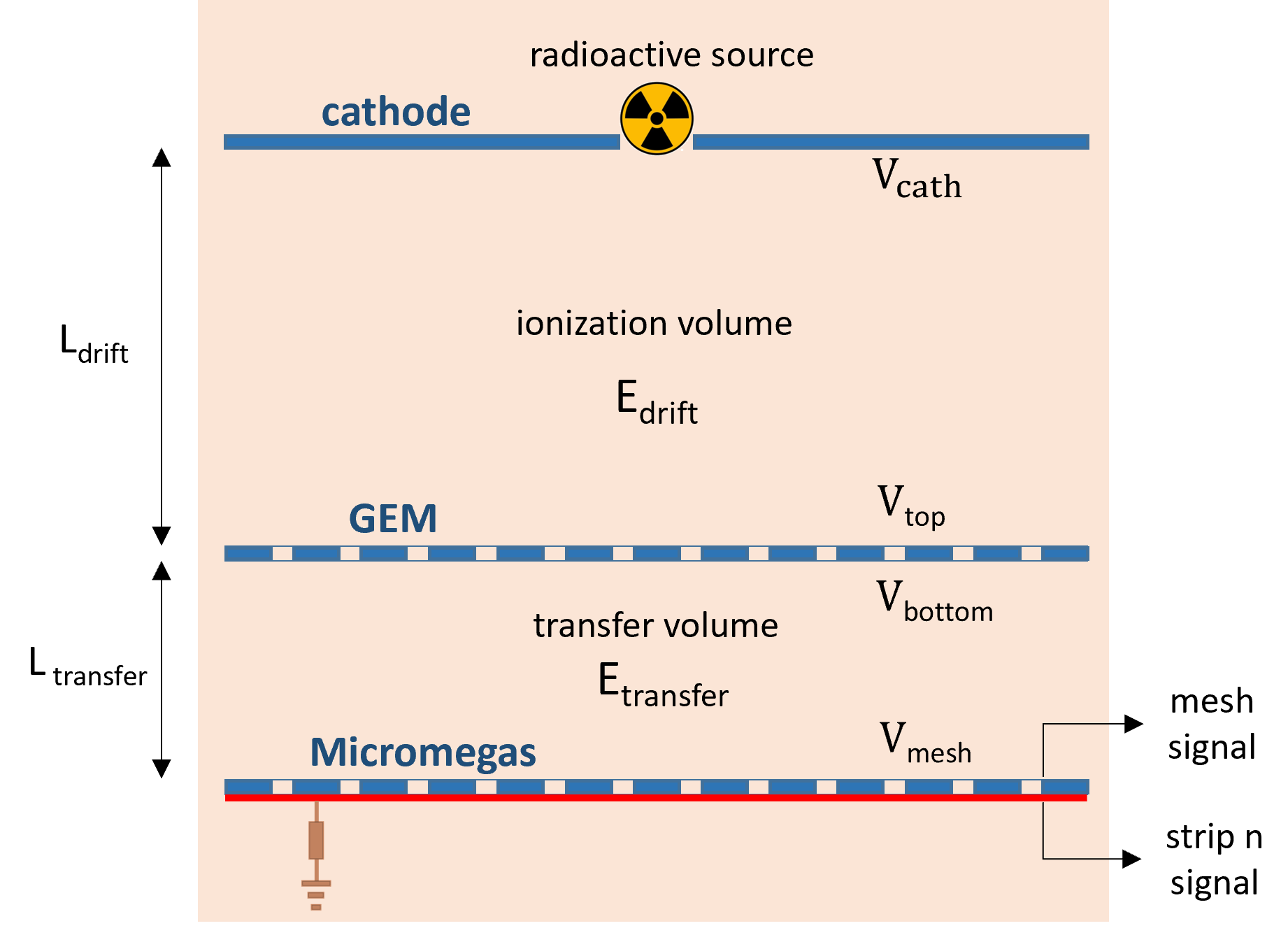}
	\caption{\label{fig:schematic_set-up} Schematic view of the different elements of the set-up (Micromegas, GEM, cathode, calibration source) along with the relevant parameters ($ L_{\mathrm{transfer}} $, $ L_{\mathrm{drift}} $, $ E_{\mathrm{transfer}} $, $ E_{\mathrm{drift}} $, $ V_{\mathrm{mesh}} $, $ V_{\mathrm{bottom}} $, $ V_{\mathrm{top}} $, $ V_{\mathrm{cath}} $).}
\end{figure}

\begin{figure}
	\centering
	\includegraphics[width=0.24\textwidth]{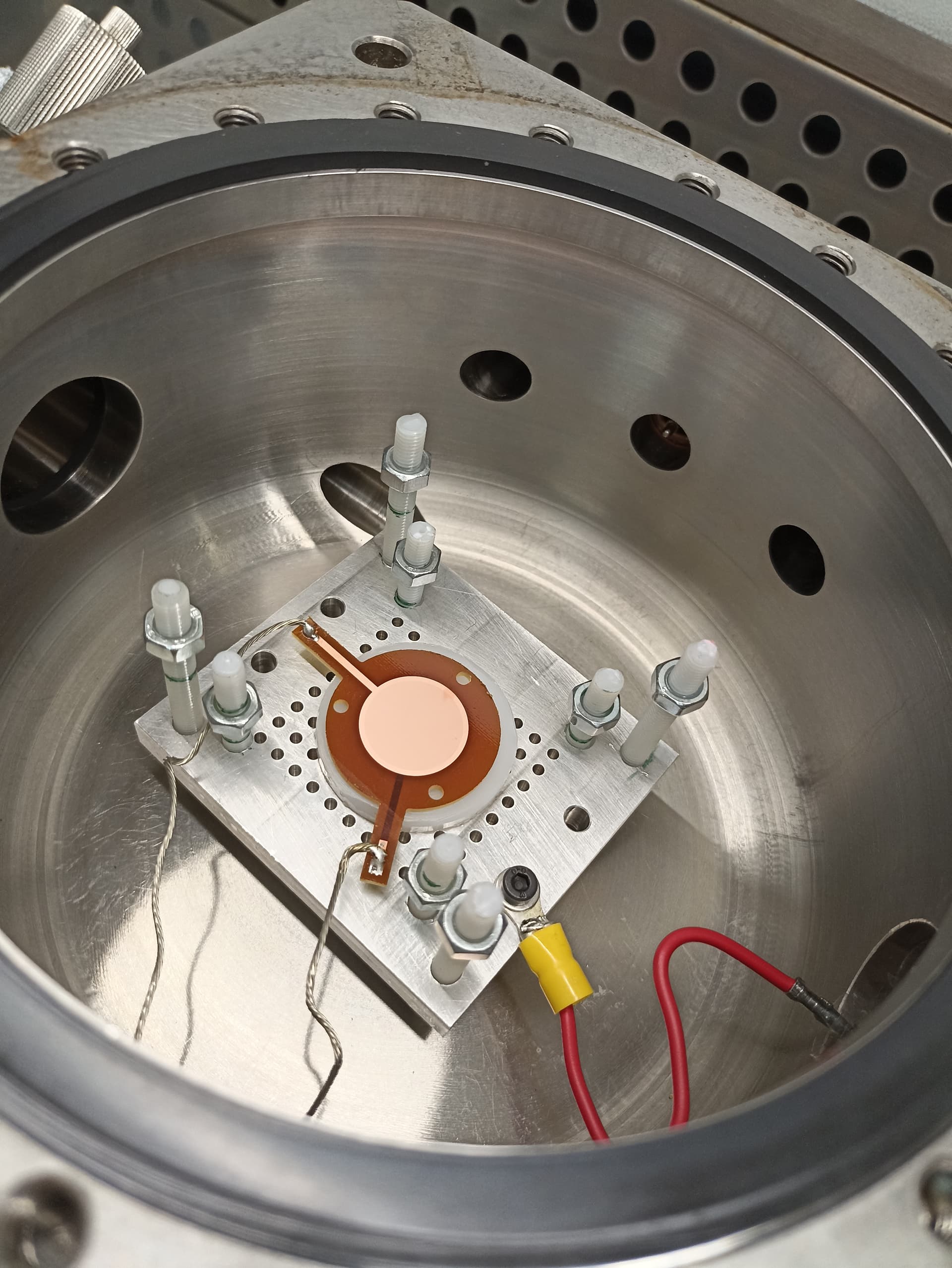}
    \includegraphics[width=0.24\textwidth]{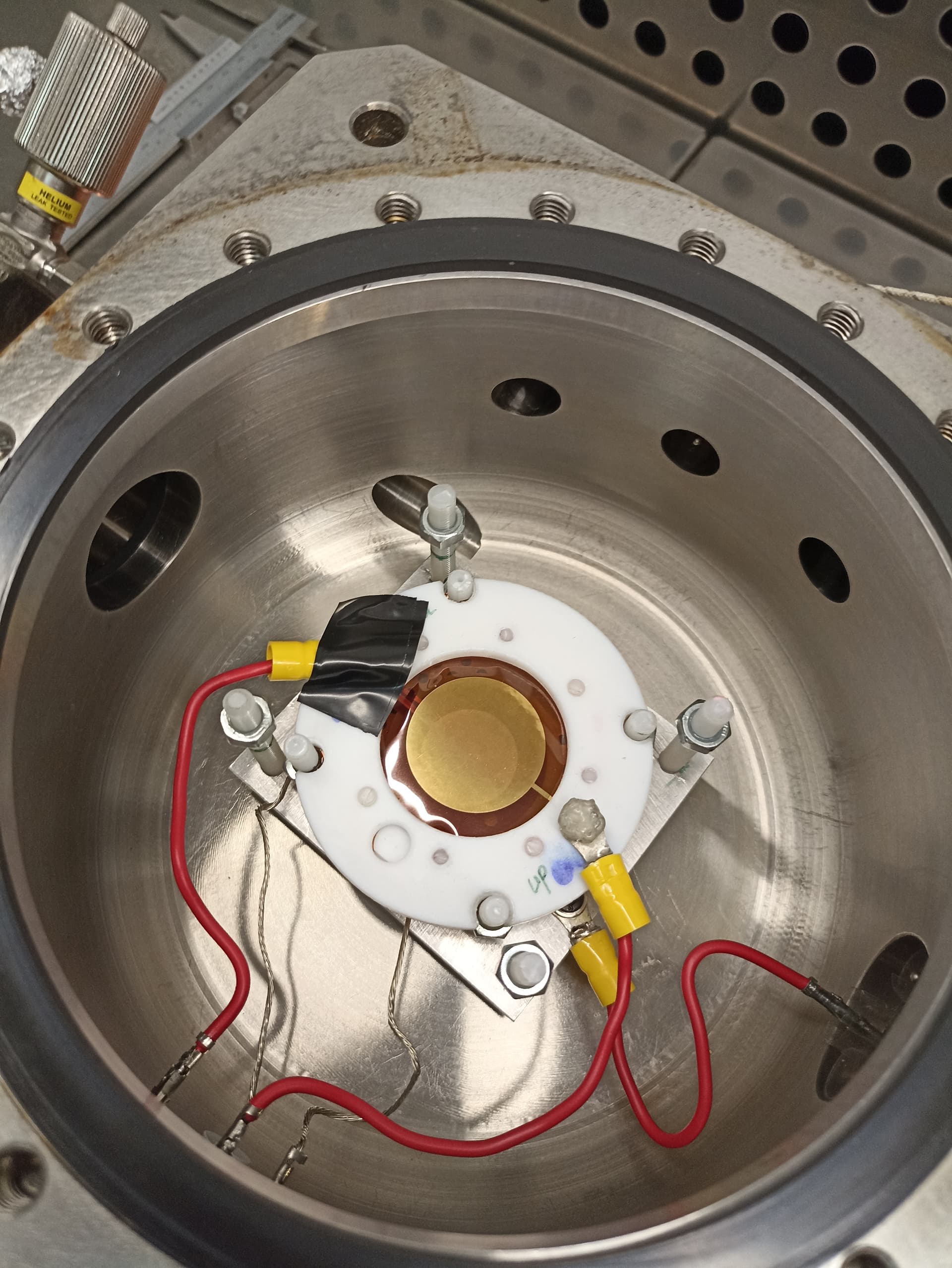}
    \includegraphics[width=0.24\textwidth]{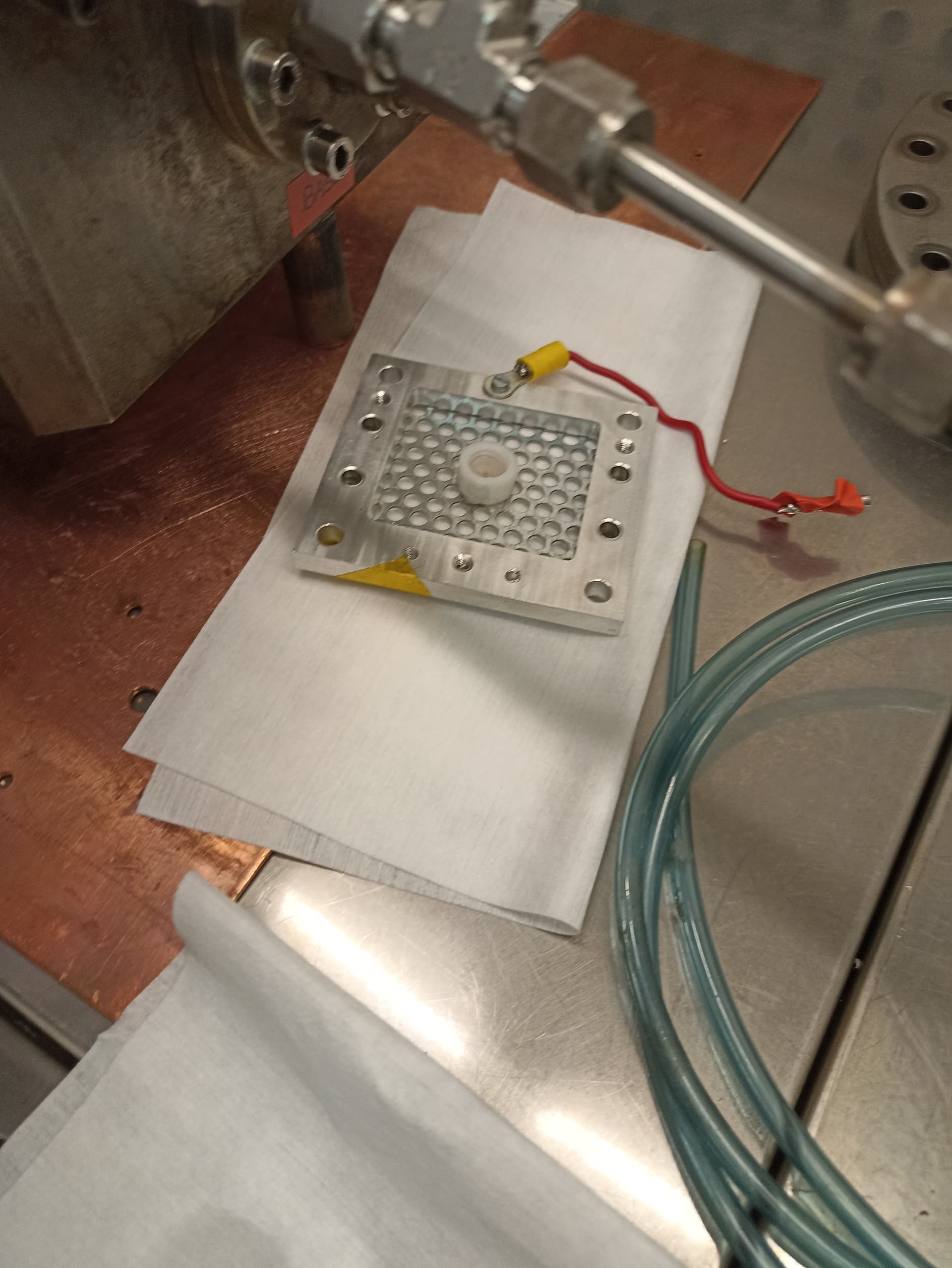}
	\caption{\label{fig:mm_gem_cath} Left: metallic support plate inside the vessel with the 2-cm-diameter Microbulk Micromegas on top. The inner PTFE pillars support the GEM at the appropriate distance, while the outer PTFE pillars are used to hold the cathode. The anode, mesh and plate connections to the feedthroughs are shown. Center: GEM foil mounted on top of the Micromegas, along with the HV connections. Right: Cathode with $ ^{55} $Fe source attached facing down.}
\end{figure}

A schematic view of the set-up can be seen in \autoref{fig:schematic_set-up}, and images of the different elements are shown in \autoref{fig:mm_gem_cath}.

Regarding the DAQ, the signal from the anode is first sent to a preamplifier (Canberra Model 2005), and then it goes through an amplifier module (Canberra Model 2022 NIM module). Both the preamplified and the amplified signals are read with an oscilloscope (a Tektronix TDS5054). A custom-made data-taking and analysis software is used to control the oscilloscope and process the data. This software enables remote operation via Ethernet connection to the local network, automating the data acquisition process from a PC. It records the individual pulses that compose each event, which are then analyzed to extract key signal parameters such as maximum amplitude, rise time or pulse area. While this set-up has been optimized for our specific experimental configuration, we are open to providing additional technical details to researchers interested in reproducing these results.

\subsection{Description of the full-scale TREX-DM set-up}
\label{sec:description_full-scale_set-up}

As already explained in \autoref{sec:introduction}, the motivation to explore the combination GEM + MM detector is to reduce the energy threshold in the low-mass WIMPs search carried out by TREX-DM. Therefore, although the results for a small set-up that will be described in \autoref{sec:results_test_set-up} look promising, a deeper investigation is required to determine if they are also achievable in real experimental conditions (essentially, a much larger readout area and drift distance). To this end, a test bench is prepared with a stainless-steel 50~L chamber containing a spare detector identical to the ones installed in TREX-DM and a GEM foil on top. This chamber achieves a vacuum level of $ \sim 10 $~mbar using a Pfeiffer Vacuum HiCube 80 Classic Turbo Pump during $ \sim 1$~h, then it is filled with Ar-1\%iC$_{4}$H$_{10}$, and a flow of 8~L h\textsuperscript{-1} is set during 72~h to ensure good quality of the gas. During these tests, the pressure has been set to 1~bar due to design specifications of the chamber. The Microbulk Micromegas detector lies on top of the endcap of the chamber. As already mentioned, the readout plane has a 25x25~cm$^2$ square active area, patterned with 512 strips (256 in each direction) with the mesh gap being 50~$ \upmu $m. A GEM foil of the same dimensions and gap is placed above the mesh, at a distance $ L_{\mathrm{transfer}} = $ 10~mm. Finally, the cathode (a stainless-steel grid) is placed above the GEM, at a distance $ L_{\mathrm{drift}} = $ 100~mm. The cathode has two $ ^{109} $Cd radioactive sources attached (K-alpha X-ray at 22.1~keV).

Again, as described in \autoref{sec:description_test_set-up}, the mesh is kept at $ V_{\mathrm{mesh}} $, the bottom and top layers of the GEM at $ V_{\mathrm{bottom}} $ and $ V_{\mathrm{top}} $, with $ V_{\mathrm{GEM}} = V_{\mathrm{top}}-V_{\mathrm{bottom}} $, and the cathode at $ V_{\mathrm{cath}} $. The same CAEN HV power supply N1471H is used in this test.

To read out the signals from the TPC, a combination composed of a Front-End Card (FEC) with AGET (ASIC for General Electronic readout of TPCs) chips~\cite{AGET:2011} and a Feminos card~\cite{FEMINOS:2014} is used. Both the FEC and the Feminos are custom-made electronics cards developed by CEA Saclay as a solution for data acquisition in nuclear and high-energy physics experiments. Each FEC has 4 AGETs, each of them with 64 channels, which makes them ideal for high-granularity readouts such as the ones used in TREX-DM. The AGETs generate the trigger based on the signal (it can be fine-tuned to trigger on single-channel pulses), and they provide the amplification, shaping and storage of the analog signals. On the other hand, the Feminos interfaces with the AGETs to digitize the analog signals (with 12-bit precision) and aggregate them into coherent events. This allows for high-speed data transfer from the Feminos towards the back-end DAQ, namely a computer using a custom-made software to interface with the Feminos card and handle the data acquisition and storage into .aqs files (raw data files). To process and analyze the data, an analysis routine based on REST-for-Physics~\cite{REST:2022}, a custom-made software framework developed for the analysis of data from rare event search experiments, is implemented.

\begin{figure}
	\centering
	\includegraphics[width=0.30\textwidth]{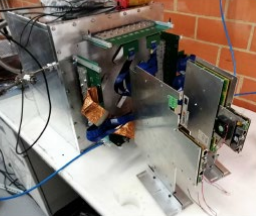}
    \includegraphics[width=0.30\textwidth]{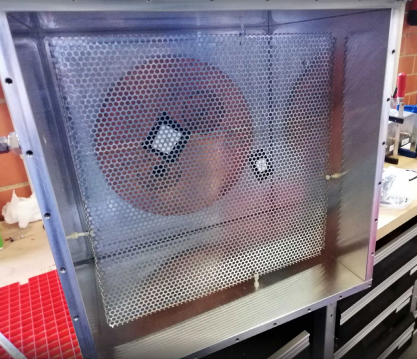}\\
    \includegraphics[width=0.30\textwidth]{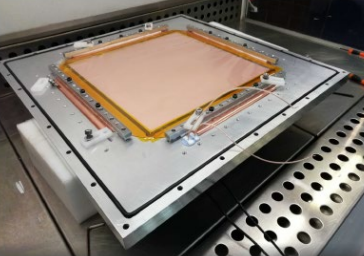}
    \includegraphics[width=0.30\textwidth]{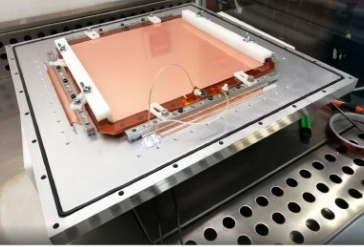}
	\caption{\label{fig:big_chamber_mm_gem_cath} Top left: closed vessel with DAQ. Top right: cathode with two $ ^{109} $Cd radioactive sources (attached with black tape). Bottom left: Micromegas detector secured on the endcap of the chamber. Bottom right: GEM foil placed on top of the mesh.}
\end{figure}

The schematic view of the set-up is shown in \autoref{fig:schematic_set-up}, and images of the assembly of all the parts of the set-up can be seen in \autoref{fig:big_chamber_mm_gem_cath}.

\section{Results}
\label{sec:results}

This section contains the main results derived from the operation of the test set-up and the full-scale setup described in \autoref{sec:methods}.

\subsection{Results from the test set-up}
\label{sec:results_test_set-up}

The goal is to obtain the relative amplification factor provided by the extra GEM stage with respect to only-Micromegas runs. To this end, both only-Micromegas ($ V_{\mathrm{mesh}} $ ON, $ V_{\mathrm{GEM}} = 0 $) and Micromegas+GEM ($ V_{\mathrm{mesh}} $ and $ V_{\mathrm{GEM}} $ ON) calibration runs are taken. We define the preamplification factor as the extra gain added by the GEM relative to a fixed Micromegas-induced gain. This is the natural extra gain parameter that arises when adding a second amplifying stage. However, in real experimental conditions, the maximum voltage that can be reached with the Micromegas alone is higher than the one that can be achieved with the GEM stage. Therefore, we define the GEM effective extra gain factor, GEM extra factor for short, as the amplification provided by the GEM in the optimized GEM + MM set-up with respect to the optimized only-Micromegas set-up. The chosen gas mixture for this study is Ar-1\%iC$_{4}$H$_{10}$, due to its importance both for IAXO~\cite{babyIAXO:2020} and TREX-DM, the main experimental pursuits of the authors, and its immediate availability. Both the preamplification factor and the GEM extra factor are obtained at 1, 4 and 10~bar (target pressure in TREX-DM). The only missing value is the preamplification factor at 1 bar, due to noise problems in the set-up at that pressure the day when those data were taken. These noise issues were traced back to fluctuations in the preamplifier and amplifier modules. Despite attempts at mitigation, including substituting alternative preamplifier and amplifier modules, we were unable to sufficiently reduce the noise. However, since the relevant parameter and focus of our efforts, the GEM extra factor, had already been measured, we proceeded with the rest of the measurements. We note that further work on noise reduction for this set-up is ongoing.

\begin{figure}
	\centering
	\includegraphics[width=0.81\textwidth]{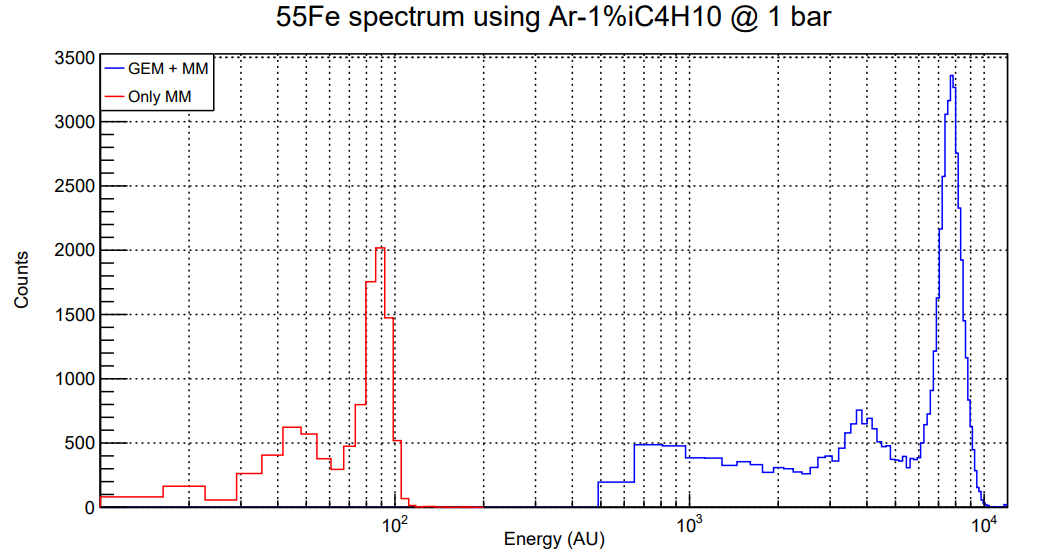}
    \includegraphics[width=0.81\textwidth]{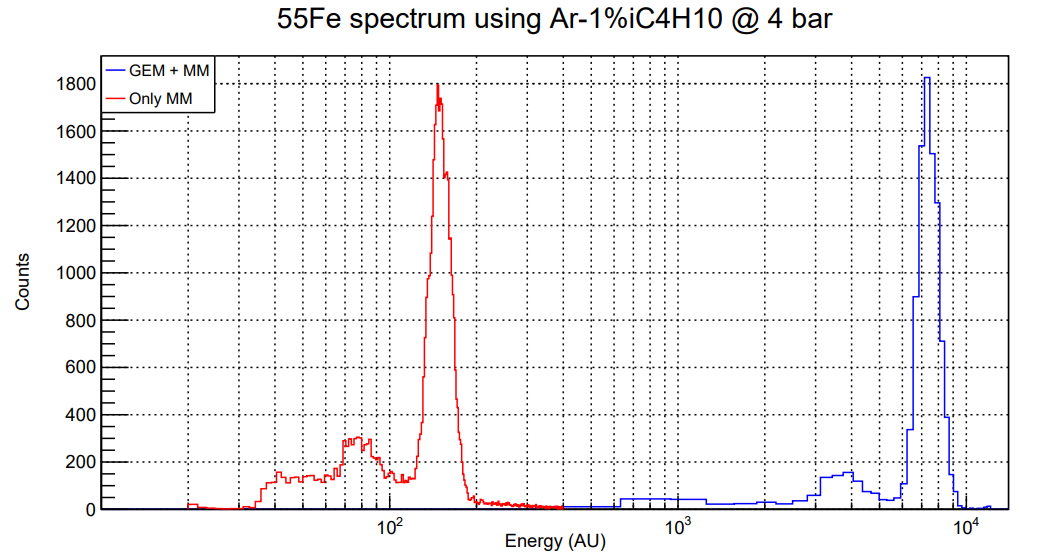}
    \includegraphics[width=0.81\textwidth]{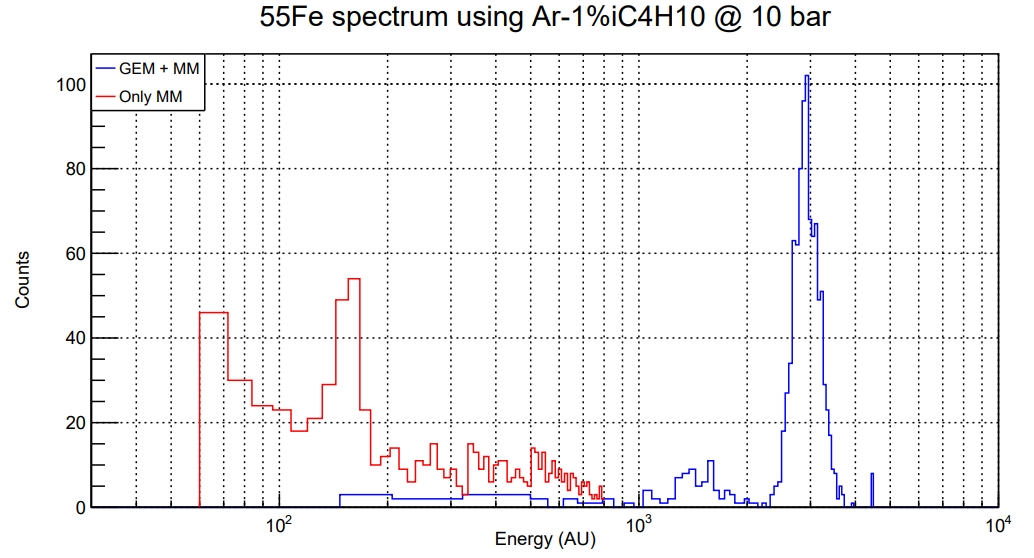}
	\caption{\label{fig:spectra_gem_mm_small} Energy spectra comparison between only MM (red) and GEM + MM (blue) calibrations using a $ ^{55} $Fe source in the test set-up. The gas mixture is Ar-1\%iC$_{4}$H$_{10}$. Note that the horizontal axis is presented in logarithmic scale. Top: 1~bar, GEM extra factor $ \approx 90 $; middle: 4~bar, GEM extra factor $ \approx 50 $; bottom: 10~bar, GEM extra factor $ \approx 20 $. The voltages of these runs are the ones recorded in the fifth column (red lines) and second and third columns (blue lines) of \autoref{table:gem_mm_data}.}
\end{figure}

The comparison of both GEM + MM and only-Micromegas spectra is done within the same dynamical range of a given electronics setup, to avoid systematics derived from different electronic gains. Thus, one must be careful selecting the parameters of the DAQ electronics, as it is easy to saturate the amplifier module with the GEM-preamplified signals. In this way, direct comparison of energy spectra like the ones shown in \autoref{fig:spectra_gem_mm_small} can be made.

The operation points with highest stable voltages achieved for the set-up described in \autoref{sec:description_test_set-up} are summarised in \autoref{table:gem_mm_data}. A reference value for the maximum voltage for only $V_{\mathrm{mesh}}$ runs was obtained from~\cite{TREXDM_Bckg_Assessment:2018}. For combined runs, the starting point for both $V_{\mathrm{mesh}}$ and $ V_{\mathrm{GEM}} $ was a safe value, around 30-40~V below the reference voltage. From that value, the voltage was raised little by little, in 5~V increments, first in $V_{\mathrm{mesh}}$ and then in $ V_{\mathrm{GEM}} $, until reaching unstable behaviour (generally sparks). It was observed, however, that for the last stable voltage in $V_{\mathrm{mesh}}$, $ V_{\mathrm{GEM}} $ could still be pushed a bit further up.

\begin{table}
	\centering
	\begin{tabledata}{$c^c^c^c^c^c} 
		\header Pressure & $ V_{\mathrm{mesh}} $ (V) & $ V_{\mathrm{GEM}} $ (V) & Preamp. & $ V_{\mathrm{mesh}} $ (V) & GEM effective \\
        \header (bar) & (GEM + MM system) & (GEM + MM system) & factor & (only-MM system) & extra gain factor \\
        \rowcolor{white}\row 1 & 305 & 310 & - & 315 & 90 \\ 
        \rowcolor{tableShading}\row 4 & 390 & 410 & 70 & 400 & 50 \\
        \rowcolor{white}\row 10 & 535 & 550 & 21 & 540 & 19 \\
        \rowcolor{tableShading}\row 1 & 290 & 285 & 85 & 293 & 80 \\ 
	\end{tabledata}
    \caption{\label{table:gem_mm_data}GEM extra factors and preamplification factors achieved for the two set-ups at different pressures, defined as the gain ratio between GEM + MM runs ($V_{\mathrm{GEM}} \neq$ 0~V) and only Micromegas runs ($ V_{\mathrm{GEM}} $= 0~V). The first three entries correspond to the test set-up, while the fourth line belongs to the full-scale set-up.}
\end{table}

Here, in all cases, $ E_{\mathrm{drift}}= 100 $~V cm\textsuperscript{-1}~bar\textsuperscript{-1} and $ E_{\mathrm{transfer}}= 100 $~V cm\textsuperscript{-1} bar\textsuperscript{-1}, values which usually lie within the electron transmission plateau of the Micromegas. However, border effects cannot be excluded given the size of the active areas and the absence of a field shaper, so these values might not represent the optimal operating conditions. Nevertheless, given that the purpose of this set-up was just to prove the feasibility and potential of the combined GEM + MM system when installed in real experimental conditions in TREX-DM, detailed field optimizations were not the focus at this stage, and therefore, the in-depth study about the electron transmission (and gain curves) of the GEM + MM was left for the full-scale set-up in \autoref{sec:results_full-scale_set-up}.

Comparison of the position of the 5.9~keV peak in $ ^{55} $Fe calibration runs points to maximum GEM extra factors of 90 (1~bar), 50 (4~bar) and 20 (10~bar). These factors have been reproduced over several days and for different Micromegas detectors with the same micropattern, and all of them are contained within a range of $ \pm 20 $\%. This variation inherently represents the systematic uncertainty in our measurements, mainly arising from slight differences in the highest stable voltages achieved (usually $ \pm $ 5~V either in $ V_{\mathrm{mesh}} $, $ V_{\mathrm{GEM}} $ or both), as well as environmental fluctuations in temperature and pressure during the measurement period. The decrease in gain with pressure is expected in MPGDs in general (even though Microbulk Micromegas do not display such a pronounced performance degradation at high pressures)~\cite{Iguaz:2022}. However, a more comprehensive theoretical analysis comparing these observed gain trends with expectations from existing models and simulations is left for future work.

Several examples of this comparison are shown in \autoref{fig:spectra_gem_mm_small}. In the case of 10~bar, the only-Micromegas runs are more difficult to take, because the mean free path of 5.9~keV photons in Ar-1\%iC$_{4}$H$_{10}$ at that pressure is 2.3~mm~\cite{NIST_XCOM:2010}, and those that do not get absorbed in the drift volume have to go through the GEM foil. Therefore, the exponential background is already noticeable and the calibration peak is less intense, but still clearly visible. All the resolutions (in \%FWHM), with and without preamplification, are around 20\%, except for the only-Micromegas run at 10 bar, which is around 30\%, mainly due to the problem with the small number of events mentioned above. However, this result points to no significant degradation in resolution when adding a preamplification stage.
 
\subsection{Results from the full-scale TREX-DM set-up}
\label{sec:results_full-scale_set-up}

Although the initial goal was to replicate the results from the small set-up at 1 bar discussed in \autoref{sec:results_test_set-up}, some more tests are performed in this full-scale set-up.

\begin{figure}
	\centering
	\includegraphics[width=0.48\textwidth]{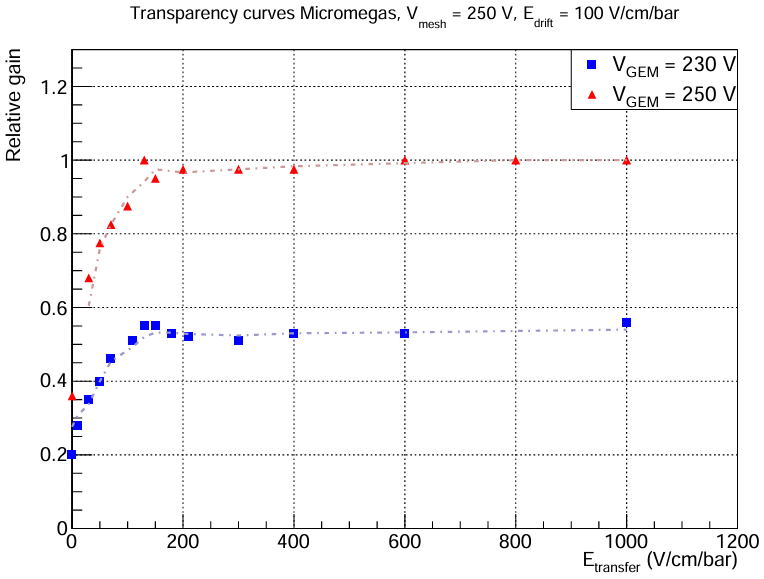}
    \includegraphics[width=0.48\textwidth]{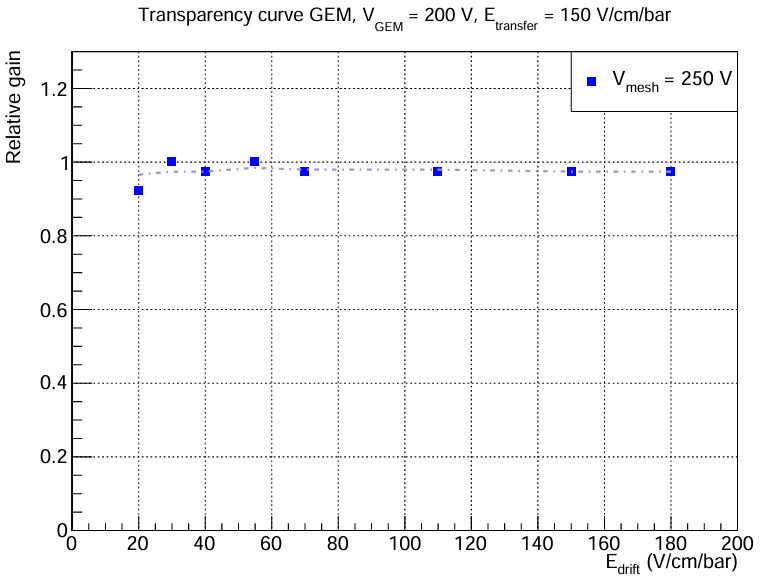}
	\caption{\label{fig:transparency_curves_full_scale} Electron transmission curves. The $y$ axis corresponds to the mean peak position, normalized to the highest value. Statistical errors in both plots are < 1\%. Left: Micromegas transmission for fixed mesh voltage and two different GEM voltages. $E_{\mathrm{drift}}$ = 100~V cm\textsuperscript{-1} bar\textsuperscript{-1} because there is total transparency in the GEM for that drift field. Right: GEM transmission for fixed mesh and GEM voltages. $E_{\mathrm{transfer}}$ = 150~V cm\textsuperscript{-1} bar\textsuperscript{-1} in order to be at the plateau of the Micromegas transparency curve.}
\end{figure}

In particular, the electron transmission (transparency curves) of the GEM foil and the mesh is studied. The results are shown in \autoref{fig:transparency_curves_full_scale}. As for the GEM transmission, it can be seen that a plateau is reached very quickly, even for very low drift field values. This is expected, because in previous studies of electron transmission in GEMs~\cite{Sauli:2002}, it has been shown that collection efficiency increases with $ V_{\mathrm{GEM}} $, meaning that full transparency is achieved with lower $E_{\mathrm{drift}}$ as $ V_{\mathrm{GEM}} $ goes up. Regarding the Micromegas curves, an interesting phenomenon occurs at $ E_{\mathrm{transfer}} = $ 0~V cm\textsuperscript{-1} bar\textsuperscript{-1}, because the relative gain has a non-zero value: the photons converted in the drift volume are amplified through the GEM, and thanks to the proximity to the mesh, diffusion is enough for some of the events to reach the Micromegas, where they are amplified again. This effect is possibly explained by the fact that $L_{\mathrm{transfer}}$ is small, $\sim$ 1~cm, but tests at different transfer distances would be necessary to shed light on this. On the other hand, the relative gain remains roughly constant from $ E_{\mathrm{transfer}} \approx $ 150~V cm\textsuperscript{-1} bar\textsuperscript{-1}. This result is unanticipated, because the transparency curve is expected to depend on the extraction efficiency of the GEM bottom layer and the collection efficiency of the Micromegas. While this plateau is usual for the Micromegas collection efficiency, one would expect the extraction efficiency of the GEM to increase in this region, up to a few kV cm\textsuperscript{-1} bar\textsuperscript{-1}~\cite{GEM:2003}. We do not have a conclusive explanation for this, but several hypotheses are being considered: 

\begin{itemize}
    \item What we think is a plateau is a slowly rising curve. Due to voltage limitations of the set-up, it is not possible to go further beyond $\sim$ 1~kV cm\textsuperscript{-1} bar\textsuperscript{-1}, so perhaps there is a gain increase up to a few kV cm\textsuperscript{-1} bar\textsuperscript{-1}.
    \item The gas mixture is playing an important role in the shape of the curves. Normally, GEM detectors use noble gases in combination with gases such as CH$_{4}$ or CF$_{4}$ due to their high drift velocities and low diffusion coefficients with respect to other quenchers such as iC$_{4}$H$_{10}$~\cite{diffusion:1984}. To the best of our knowledge, the mixture Ar-1\%iC$_{4}$H$_{10}$ has not been characterised in the context of GEMs, and it could be the case that its higher diffusion coefficients imply a loss of extraction efficiency from the bottom layer of the GEM.
\end{itemize}

Irrespective of the reason, it should be noted that this unexplained behaviour does not invalidate the results: at most, an optimisation of the extraction efficiency would yield higher gains, and thus larger GEM extra factors.

\begin{figure}
	\centering
	\includegraphics[width=0.48\textwidth]{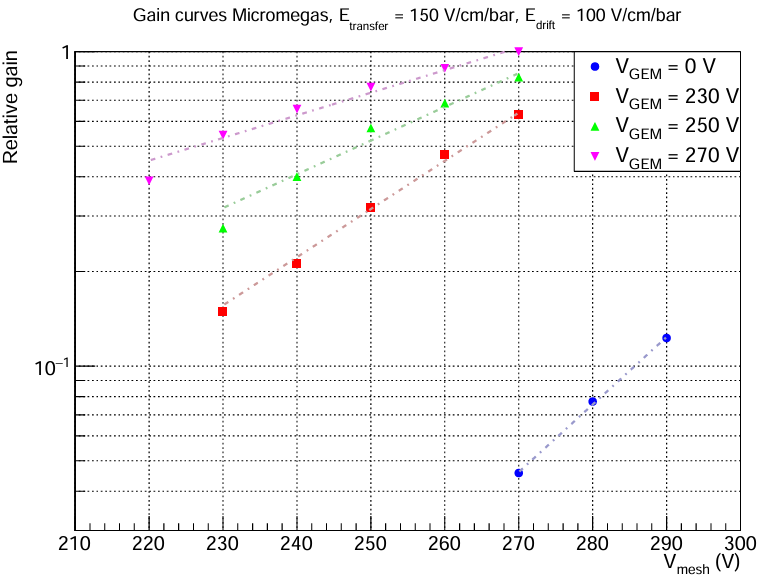}
    \includegraphics[width=0.48\textwidth]{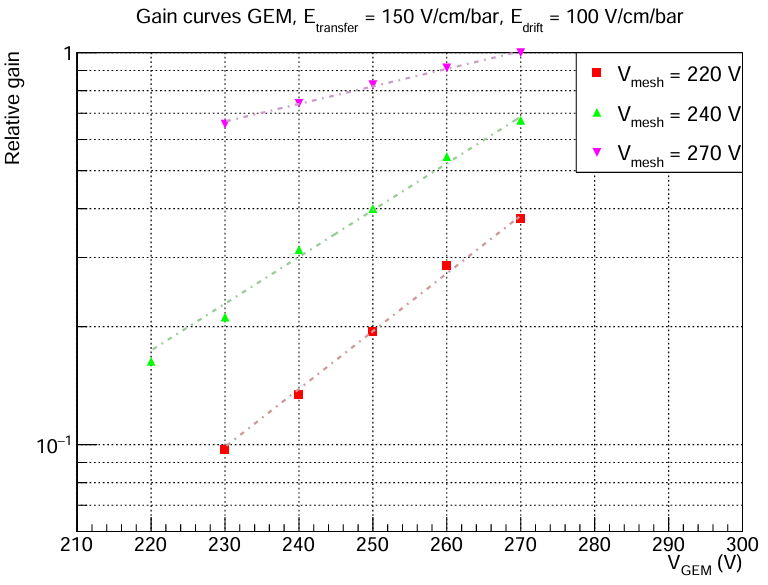}
	\caption{\label{fig:gain_curves_full_scale} Gain curves. The $y$ axis corresponds to the mean peak position, normalized to the highest value. Statistical errors in both plots are < 1\%. Left: Micromegas curves for fixed $ V_{\mathrm{GEM}} $. Right: GEM curves for fixed $ V_{\mathrm{mesh}} $. In both cases, $E_{\mathrm{drift}}$ and $E_{\mathrm{transfer}}$ are fixed. Note that the maximum gain corresponds to the same data point in both plots ($ V_{\mathrm{mesh}}= 270 $~V, $ V_{\mathrm{GEM}} = 270 $~V), so the relative gain is directly comparable between plots.}
\end{figure}

Also, gain curves are studied before the determination of the maximum GEM extra factor. As it can be seen in \autoref{fig:gain_curves_full_scale}, two types of gain curves are examined: Micromegas gain curves (varying $ V_{\mathrm{mesh}} $ for a fixed $ V_{\mathrm{GEM}} $) and GEM gain curves (varying $ V_{\mathrm{GEM}} $ for a fixed $ V_{\mathrm{mesh}} $). In both cases, the expected exponential behaviour with amplification voltage is observed. In the Micromegas curves, the special case $ V_{\mathrm{GEM}} $ = 0~V is also included, which corresponds to the baseline only-Micromegas detector. Comparison of this curve with the GEM + MM curves already hints at extra gain factors of O(10) thanks to the GEM addition. In all cases, it can be seen that the curves are not perfectly parallel (in log scale). This suggests both stages are not totally factorizable, possibly due to ion backflow or other complex interactions between the GEM and Micromegas stages. To investigate this effect further, a dedicated set-up based on a small chamber similar to the one presented in \autoref{sec:description_test_set-up} is being prepared in Zaragoza. In this set-up, we will perform systematic studies using UV laser calibration techniques with a photocathode. Furthermore, measurements with the GEM positioned at different distances relative to the Micromegas are planned to better understand these interactions. While a more detailed analysis will be the subject of future work, it is not critical for the primary objectives of this paper.

\begin{figure}
	\centering
	\includegraphics[width=0.9\textwidth]{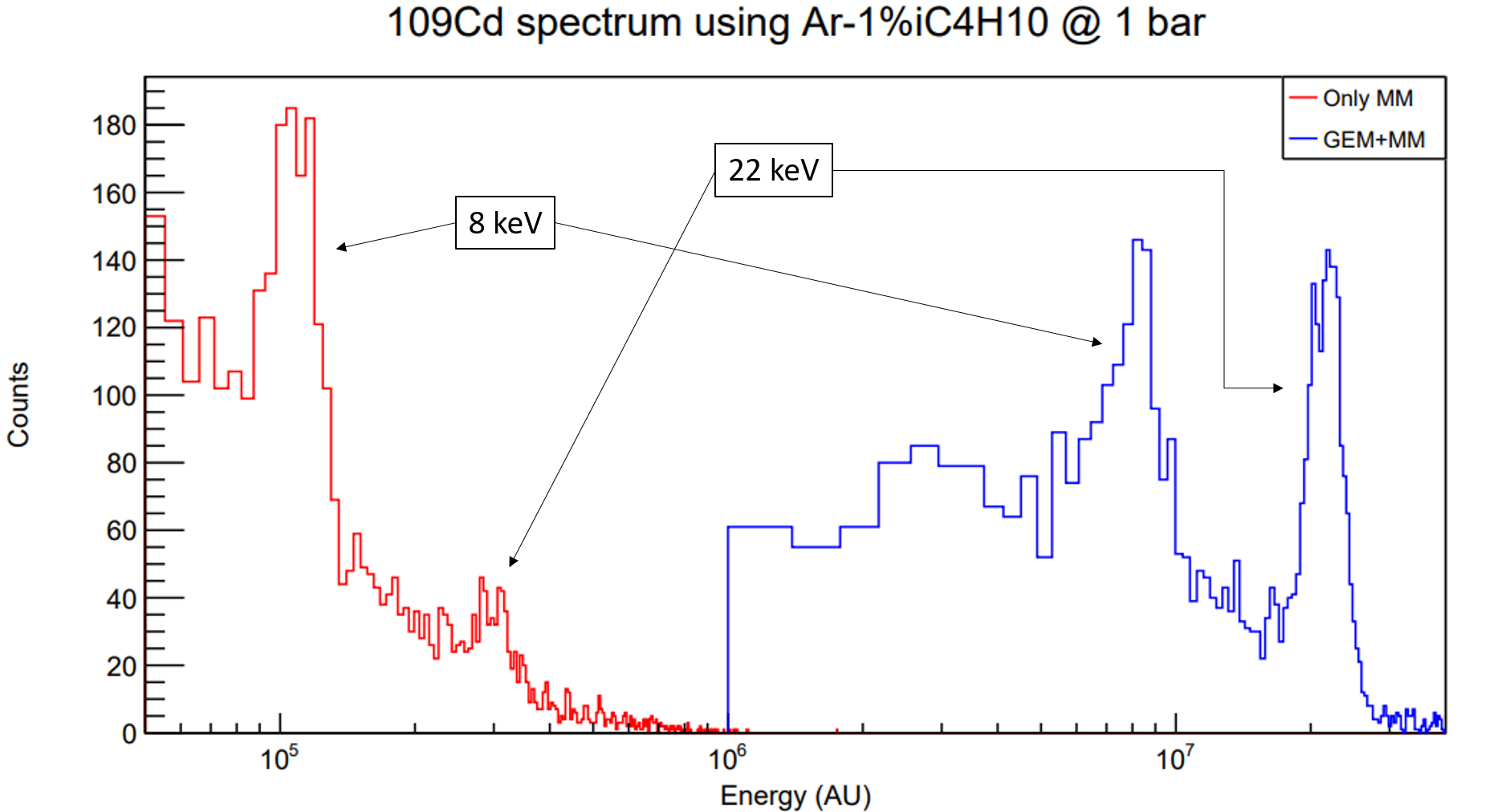}
	\caption{\label{fig:1bar_preamp_factor_trex_gem} Energy spectrum comparison between only MM and GEM + MM calibrations using a $ ^{109} $Cd source in the full-scale set-up. The 8~keV peak corresponds to the copper fluorescence at the Micromegas surface. The gas mixture is Ar-1\%iC$_{4}$H$_{10}$ at 1~bar. Note the horizontal axis is presented in logarithmic scale, and that an energy cut has been applied to the GEM + MM run in order to remove background and keep the left part of the canvas clean.}
\end{figure}

Lastly, some more data are taken in order to explore the value for the maximum GEM extra factor and preamplification factor achievable. In the fourth row of \autoref{table:gem_mm_data}, the achieved voltages are shown. Comparison of the position of the 8~keV copper fluorescence peak in $ ^{109} $Cd calibration runs (see \autoref{fig:1bar_preamp_factor_trex_gem}) yields a GEM extra factor of 80, in line with the result obtained in \autoref{sec:results_test_set-up}. In these runs, $ E_{\mathrm{drift}}= 100 $~V cm\textsuperscript{-1} bar\textsuperscript{-1}, $ E_{\mathrm{transfer}}= 150 $~V cm\textsuperscript{-1} bar\textsuperscript{-1} as suggested by the transparency curves.

Note that the voltages are lower than those presented in \autoref{sec:results_test_set-up} because of the intrinsic difficulty associated to operating larger-area Micromegas (1 vs. 1024 channels means higher possibility of leakage currents between mesh and some channels). To address this challenge, we are investigating the implementation of resistive Micromegas technology~\cite{Attie:2013}, which has shown improved stability against spark formation in similar applications. This approach could potentially allow operation at higher voltages in the full-scale detector, bringing performance closer to that observed in the small-scale tests. Despite the current voltage limitations, the potential to lower the energy threshold is still present even in the full-scale set-up, mimicking the real experimental conditions of TREX-DM. 

\section{Discussion}
\label{sec:discussion}

This section discusses the impact the results presented in \autoref{sec:results} have on the sensitivity of TREX-DM. As argued in the introduction, there is a strong motivation to extend the sensitivity of dark matter experiments to lower WIMP masses. This requires lowering energy thresholds, as well as reducing the background at the lowest energies.   

As of 2022, TREX-DM achieved a background level at low energies of around 80~dru (dru = c keV\textsuperscript{-1} day\textsuperscript{-1} kg\textsuperscript{-1}), and an energy threshold around 900~eV$_{ee}$. Reducing the threshold from 900~eV$_{ee}$ to 50~eV$_{ee}$ significantly enhances sensitivity in the < 1~GeV c\textsuperscript{-2} region. As already demonstrated in this paper, a great improvement in energy threshold can be achieved by introducing a new electron preamplification stage (the GEM) atop the Micromegas. Laboratory tests (\autoref{sec:results_test_set-up} and~\autoref{sec:results_full-scale_set-up}) have proven GEM extra factors ranging from 20 to 90 are feasible, contingent on gas pressure.

\begin{figure}
	\centering
	\includegraphics[width=0.85\textwidth]{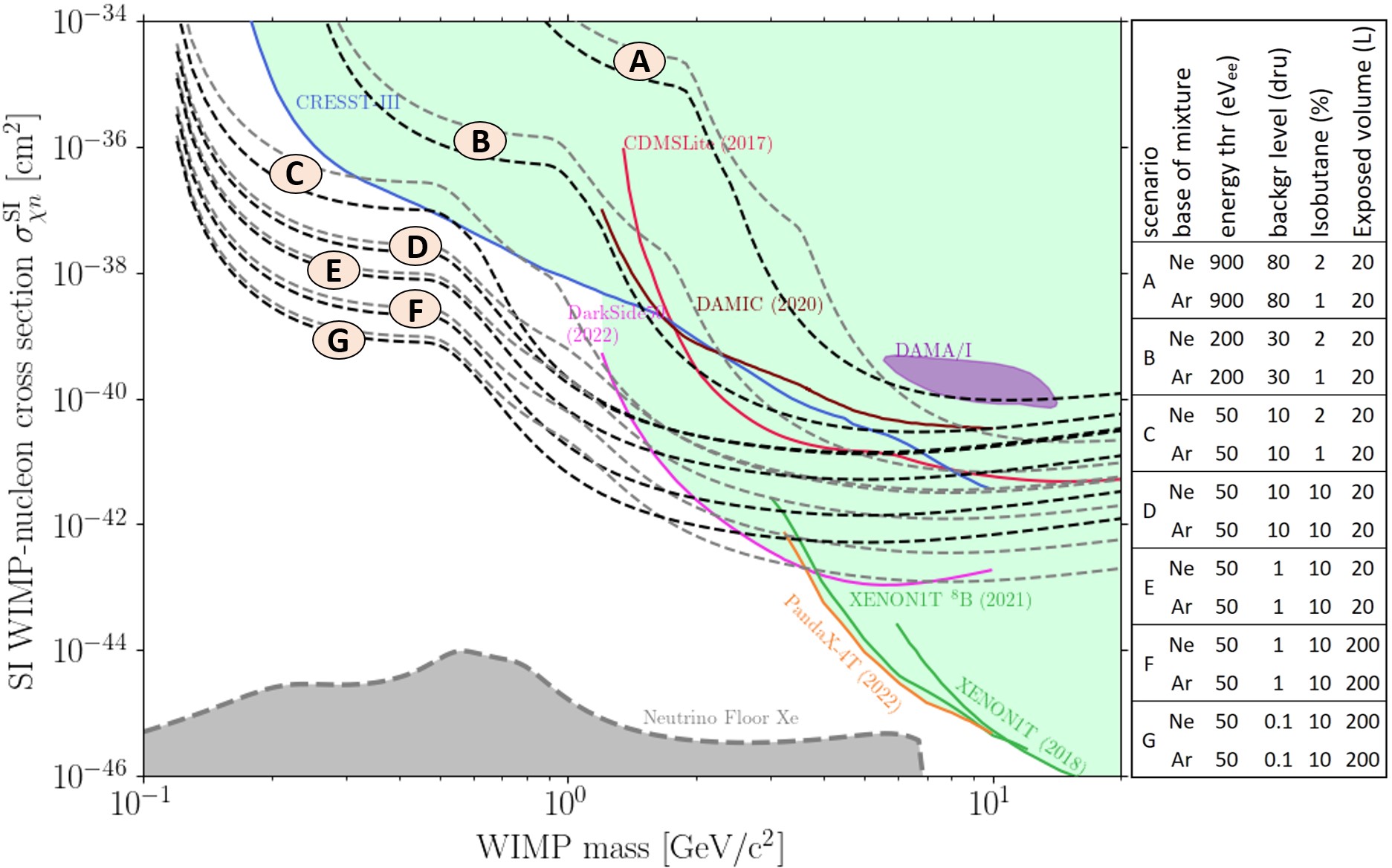}
	\caption{\label{fig:TREXDMWIMPexclusionPlot} WIMP-nucleon cross-section vs. WIMP mass exclusion plot, with current bounds from experiments~\cite{CDMSLite:2018,XENON1T:2018,CREST-III:2019,DAMIC:2020,XENON1T:2021,DarkSide50:2022,PandaX-4T:2022}, claimed discovery~\cite{DAMA/I:2013} and different scenarios for TREX-DM (all of them 1~year of exposure time). Each scenario is plotted
    with Ne-based (black) and Ar-based (grey) mixtures. Also, solar neutrino floor with a Xe target is shown~\cite{Neutrino_Floor_Xe:2014}.}
\end{figure}

In \autoref{fig:TREXDMWIMPexclusionPlot}, the sensitivity projections in TREX-DM for Spin Independent WIMP-nucleon interaction over a year are shown, considering several experimental parameters: energy threshold, background level and isobutane content of the gas mixture. The background level is expected to improve in the near term (there is a roadmap of upgrades underway, but the details are beyond the scope of this paper). On the other hand, optimizing the gas mixture plays a pivotal role. Sensitivity estimates indicate better performance with neon-based mixtures compared to argon. Also, argon mixtures with increased isobutane content are being considered: more isobutane enhances sensitivity to WIMPs below 1~GeV c\textsuperscript{-2} due to the lower mass of target nuclei. Even though these mixtures have not been studied in this paper, the preamplification results shown here are expected to hold because Ar-1\%iC$_{4}$H$_{10}$ is a conservative mixture choice: neon-based mixtures typically provide higher gains~\cite{TREXDM_Bckg_Assessment:2018}, and an increased isobutane content also goes in the direction of attaining greater amplifications. 

\section{Conclusions}
\label{sec:conclusions}

Electron amplification in gas offers an attractive strategy to increase signal-to-noise ratio and therefore to reduce detector energy thresholds. This feature, coupled with the ability of building radiopure readout Micromegas planes with the Microbulk technology, is at the core of the TREX-DM proposal. In this paper, we have demonstrated the feasibility of an amplification scheme that should allow to approach the single-electron sensitivity in realistic TREX-DM implementations. The addition of a GEM preamplification stage on top of the Microbulk readout allows for this gain, without jeopardizing the radiopurity specifications of the readout. At the time of writing this article, a GEM + MM combined readout like the one tested in \autoref{sec:description_full-scale_set-up} is being installed and commissioned at the TREX-DM experiment. As discussed in \autoref{sec:discussion}, this improvement would open a new detection window at lower recoil energies that, depending on the background levels achieved at those energies, might lead to substantial improvement to low mass WIMPs, potentially down to an unexplored region of the parameter space.

\section*{Data availability}
\subsection*{Underlying data}

Zenodo: Data for manuscript submitted to Open Research Europe with title "Micromegas with GEM preamplification for enhanced energy threshold in low-background gaseous time projection chambers"\\ \url{https://doi.org/10.5281/zenodo.14525554} \\
This project contains the following underlying data:
\begin{itemize}
	\item Spectra for the test set-up:
        \begin{itemize}
            \item spectrum\_test\_set-up\_1\_bar\_only\_MM.root (1 bar, only MM)
            \item spectrum\_test\_set-up\_1\_bar\_GEM\_MM.root (1 bar, GEM+MM)
            \item spectrum\_test\_set-up\_4\_bar\_only\_MM.root (4 bar, only MM)
            \item spectrum\_test\_set-up\_4\_bar\_GEM\_MM.root (4 bar, GEM+MM)
            \item spectrum\_test\_set-up\_10\_bar\_only\_MM.root (10 bar, only MM)
            \item spectrum\_test\_set-up\_10\_bar\_GEM\_MM.root (10 bar, GEM+MM)
        \end{itemize}
	\item Spectrum for the full-scale set-up:
        \begin{itemize}
            \item spectrum\_full-scale\_set-up\_1\_bar\_only\_MM.root (1 bar, only MM)
            \item spectrum\_full-scale\_set-up\_1\_bar\_GEM\_MM.root (1 bar, GEM+MM)
        \end{itemize}
        \item Transparency curves for the full-scale set-up:
        \begin{itemize}
            \item transparency\_curve\_micromegas\_Edrift100\_Vmesh250\_Vgem230\_full-scale\_set-up.txt 
            \item transparency\_curve\_micromegas\_Edrift100\_Vmesh250\_Vgem250\_full-scale\_set-up.txt
            \item transparency\_curve\_gem\_Etransfer150\_Vmesh250\_Vgem200\_full-scale\_set-up.txt
        \end{itemize}
        \item GEM gain curves for the full-scale set-up:
        \begin{itemize}
            \item gain\_curve\_gem\_Etransfer150\_Edrift100\_Vmesh220\_full-scale\_set-up.txt
            \item gain\_curve\_gem\_Etransfer150\_Edrift100\_Vmesh240\_full-scale\_set-up.txt
            \item gain\_curve\_gem\_Etransfer150\_Edrift100\_Vmesh270\_full-scale\_set-up.txt
        \end{itemize}
        \item Micromegas gain curves for the full-scale set-up:
        \begin{itemize}
            \item gain\_curve\_micromegas\_Etransfer150\_Edrift100\_Vgem0\_full-scale\_set-up.txt
            \item gain\_curve\_micromegas\_Etransfer150\_Edrift100\_Vgem230\_full-scale\_set-up.txt
            \item gain\_curve\_micromegas\_Etransfer150\_Edrift100\_Vgem250\_full-scale\_set-up.txt
            \item gain\_curve\_micromegas\_Etransfer150\_Edrift100\_Vgem270\_full-scale\_set-up.txt
        \end{itemize}
\end{itemize}

Note: the .root format is associated to ROOT, the CERN open-source data analysis framework (\url{https://root.cern/}).
Data are available under the Creative Commons Attribution 4.0 International license (CC-BY 4.0)\\ \url{https://creativecommons.org/licenses/by/4.0/}

\section*{Competing interests}
No competing interests were disclosed.

\section*{Grant information}
This project has received funding from the European Research Council (ERC) under the European Union’s Horizon 2020 research and innovation programme (Grant agreement No. 788781) (Towards the detection of the axion with the International Axion Observatory (IAXOplus)). This project has also received funding from the European Union’s Horizon 2020 research and innovation programme under the Marie Skłodowska-Curie grant agreement No. 101026819 (Low background techniques on particle detectors for Rare Event Searches (LOBRES)). This work is also associated to the IDEAS programme of the 7th EU Framework Programme, grant agreement ERC-2009-StG-240054 (TREX).
We would also like to acknowledge the support from the Agencia Estatal de Investigación (AEI) under the grant agreement PID2019-108122GB-C31 funded by MCIN/AEI/10.13039/501100011033, and under the grant agreement PID2022-137268NB-C51 funded by MCIN/AEI/10.13039/501100011033/FEDER, as well as funds from “European Union NextGenerationEU/PRTR” (Planes complementarios, Programa de Astrofísica y Física de Altas Energías). Finally, we would also like to acknowledge support from Gobierno de Aragón through their predoctoral research contracts, competitive call 2020-2024.

\section*{Acknowledgments}
We would like to thank the Servicio General de Apoyo a la Investigación-SAI, Universidad de Zaragoza, for their technical support, and the Micro-Pattern Technologies (MPT) workshop at CERN, where both the Micromegas and the GEMs used in this article were manufactured.

This article has also been published on arXiv. It is available at \url{https://doi.org/10.48550/arXiv.2412.19864}.

{\small\bibliographystyle{unsrtnat}
\bibliography{references}}

\begin{thebibliography}{41}
\providecommand{\natexlab}[1]{#1}
\providecommand{\url}[1]{\texttt{#1}}
\expandafter\ifx\csname urlstyle\endcsname\relax
  \providecommand{\doi}[1]{doi: #1}\else
  \providecommand{\doi}{doi: \begingroup \urlstyle{rm}\Url}\fi

\bibitem[Hilke(2010)]{Hilke:2010zz}
H.~J. Hilke.
\newblock {T}ime projection chambers.
\newblock \emph{Reports on Progress in Physics}, 73\penalty0 (11):\penalty0
  116201, Oct 2010.
\newblock \doi{10.1088/0034-4885/73/11/116201}.
\newblock URL \url{https://dx.doi.org/10.1088/0034-4885/73/11/116201}.

\bibitem[Giomataris et~al.(1996)Giomataris, Rebourgeard, Robert, and
  Charpak]{Giomataris:1995fq}
Y.~Giomataris, P.~Rebourgeard, J.~P. Robert, and G.~Charpak.
\newblock {MICROMEGAS}: {A} high granularity position sensitive gaseous
  detector for high particle flux environments.
\newblock \emph{Nucl. Instrum. Meth. A}, 376\penalty0 (1):\penalty0 29--35,
  1996.
\newblock \doi{10.1016/0168-9002(96)00175-1}.
\newblock URL
  \url{https://www.sciencedirect.com/science/article/pii/0168900296001751}.

\bibitem[Sauli(1997)]{Sauli:1997}
F.~Sauli.
\newblock {GEM}: {A} new concept for electron amplification in gas detectors.
\newblock \emph{Nuclear Instruments and Methods in Physics Research Section A:
  Accelerators, Spectrometers, Detectors and Associated Equipment},
  386\penalty0 (2):\penalty0 531--534, 1997.
\newblock ISSN 0168-9002.
\newblock \doi{10.1016/S0168-9002(96)01172-2}.
\newblock URL
  \url{https://www.sciencedirect.com/science/article/pii/S0168900296011722}.

\bibitem[Sauli(2014)]{Sauli:2014}
F.~Sauli.
\newblock 8.22 - {G}as {E}lectron {M}ultiplier ({GEM}) {D}etectors:
  {P}rinciples of {O}peration and {A}pplications.
\newblock In Anders Brahme, editor, \emph{Comprehensive Biomedical Physics},
  pages 367--408. Elsevier, Oxford, 2014.
\newblock ISBN 978-0-444-53633-4.
\newblock \doi{10.1016/B978-0-444-53632-7.00625-0}.
\newblock URL
  \url{https://www.sciencedirect.com/science/article/pii/B9780444536327006250}.

\bibitem[G\'omez(2019)]{Gomez:2019}
H.~G\'omez.
\newblock {M}uon tomography using {M}icromegas detectors: {F}rom {A}rchaeology
  to nuclear safety applications.
\newblock \emph{Nuclear Instruments and Methods in Physics Research Section A:
  Accelerators, Spectrometers, Detectors and Associated Equipment},
  936:\penalty0 14--17, 2019.
\newblock ISSN 0168-9002.
\newblock \doi{https://doi.org/10.1016/j.nima.2018.10.011}.
\newblock URL
  \url{https://www.sciencedirect.com/science/article/pii/S0168900218313251}.

\bibitem[L\'azaro~Roche(2021)]{Roche:2021}
I.~L\'azaro~Roche.
\newblock {A} {C}ompact {M}uon {T}racker for {D}ynamic {T}omography of
  {D}ensity {B}ased on a {T}hin {T}ime {P}rojection {C}hamber with {M}icromegas
  {R}eadout.
\newblock \emph{Particles}, 4\penalty0 (3):\penalty0 333--342, 2021.
\newblock ISSN 2571-712X.
\newblock URL \url{https://www.mdpi.com/2571-712X/4/3/28}.

\bibitem[Andriamonje et~al.(2010)Andriamonje, Attie, Berthoumieux, Calviani,
  Colas, Dafni, Fanourakis, Ferrer-Ribas, Galan, Geralis, Giganon, Giomataris,
  Gris, Sanchez, Gunsing, Iguaz, Irastorza, Oliveira, Papaevangelou, Ruz,
  Savvidis, Teixera, and Tomás]{Andriamonje:2010}
S.~Andriamonje, D.~Attie, E.~Berthoumieux, M.~Calviani, P.~Colas, T.~Dafni,
  G.~Fanourakis, E.~Ferrer-Ribas, J.~Galan, T.~Geralis, A.~Giganon,
  I.~Giomataris, A.~Gris, C.~Guerrero Sanchez, F.~Gunsing, F.~J. Iguaz,
  I.~Irastorza, R.~De Oliveira, T.~Papaevangelou, J.~Ruz, I.~Savvidis,
  A.~Teixera, and A.~Tomás.
\newblock {D}evelopment and performance of {M}icrobulk {M}icromegas detectors.
\newblock \emph{Journal of Instrumentation}, 5\penalty0 (02):\penalty0 P02001,
  Feb 2010.
\newblock \doi{10.1088/1748-0221/5/02/P02001}.
\newblock URL \url{https://dx.doi.org/10.1088/1748-0221/5/02/P02001}.

\bibitem[Cebrián et~al.(2011)Cebrián, Dafni, Ferrer-Ribas, Galán,
  Giomataris, Gómez, Iguaz, Irastorza, Luzón, {de Oliveira}, Rodríguez,
  Seguí, Tomás, and Villar]{Cebrian:2010}
S.~Cebrián, T.~Dafni, E.~Ferrer-Ribas, J.~Galán, I.~Giomataris, H.~Gómez,
  F.J. Iguaz, I.G. Irastorza, G.~Luzón, R.~{de Oliveira}, A.~Rodríguez,
  L.~Seguí, A.~Tomás, and J.A. Villar.
\newblock Radiopurity of {M}icromegas readout planes.
\newblock \emph{Astroparticle Physics}, 34\penalty0 (6):\penalty0 354--359,
  2011.
\newblock ISSN 0927-6505.
\newblock \doi{10.1016/j.astropartphys.2010.09.003}.
\newblock URL
  \url{https://www.sciencedirect.com/science/article/pii/S0927650510001805}.

\bibitem[Irastorza et~al.(2016{\natexlab{a}})Irastorza, Aznar, Castel,
  Cebrián, Dafni, Galán, Garcia, Garza, Gómez, Herrera, Iguaz, Luzon,
  Mirallas, Ruiz, Seguí, and Tomás]{T-REX_1:2016}
I.G. Irastorza, F.~Aznar, J.~Castel, S.~Cebrián, T.~Dafni, J.~Galán, J.A.
  Garcia, J.G. Garza, H.~Gómez, D.C. Herrera, F.J. Iguaz, G.~Luzon,
  H.~Mirallas, E.~Ruiz, L.~Seguí, and A.~Tomás.
\newblock {G}aseous time projection chambers for rare event detection:
  {R}esults from the {T-REX} project. {I}. {D}ouble beta decay.
\newblock \emph{Journal of Cosmology and Astroparticle Physics}, 2016\penalty0
  (01):\penalty0 033, Jan 2016{\natexlab{a}}.
\newblock \doi{10.1088/1475-7516/2016/01/033}.
\newblock URL \url{https://dx.doi.org/10.1088/1475-7516/2016/01/033}.

\bibitem[Irastorza et~al.(2016{\natexlab{b}})Irastorza, Aznar, Castel,
  Cebrián, Dafni, Galán, Garcia, Garza, Gómez, Herrera, Iguaz, Luzon,
  Mirallas, Ruiz, Seguí, and Tomás]{T-REX_2:2016}
I.G. Irastorza, F.~Aznar, J.~Castel, S.~Cebrián, T.~Dafni, J.~Galán, J.A.
  Garcia, J.G. Garza, H.~Gómez, D.C. Herrera, F.J. Iguaz, G.~Luzon,
  H.~Mirallas, E.~Ruiz, L.~Seguí, and A.~Tomás.
\newblock {G}aseous time projection chambers for rare event detection:
  {R}esults from the {T-REX} project. {II}. {D}ark matter.
\newblock \emph{Journal of Cosmology and Astroparticle Physics}, 2016\penalty0
  (01):\penalty0 034, Jan 2016{\natexlab{b}}.
\newblock \doi{10.1088/1475-7516/2016/01/034}.
\newblock URL \url{https://dx.doi.org/10.1088/1475-7516/2016/01/034}.

\bibitem[Galán et~al.(2010)Galán, Aune, Carmona, Dafni, Fanourakis, Ribas,
  Geralis, Giomataris, Gómez, Iguaz, Irastorza, Kousouris, Luzón, Morales,
  Mols, Papaevangelou, Rodríguez, Ruz, Tomás, and
  Vafeiadis]{MicrobulkCAST_2010}
J.~Galán, S.~Aune, J.~Carmona, T.~Dafni, G.~Fanourakis, E.~Ferrer Ribas,
  T.~Geralis, I.~Giomataris, H.~Gómez, F.~J. Iguaz, I.~G. Irastorza,
  K.~Kousouris, G.~Luzón, J.~Morales, J.~P. Mols, T.~Papaevangelou,
  A.~Rodríguez, J.~Ruz, A.~Tomás, and T.~Vafeiadis.
\newblock {MICROMEGAS} detectors in the {CAST} experiment.
\newblock \emph{Journal of Instrumentation}, 5\penalty0 (01):\penalty0 P01009,
  Jan 2010.
\newblock \doi{10.1088/1748-0221/5/01/P01009}.
\newblock URL \url{https://dx.doi.org/10.1088/1748-0221/5/01/P01009}.

\bibitem[Ferrer-Ribas et~al.(2023)Ferrer-Ribas, Altenmüller, Biasuzzi, Castel,
  Cebrián, Dafni, Desch, Díez-Ibañez, Galán, Galindo, García, Giganon,
  Goblin, Irastorza, Kaminski, Luzón, Margalejo, Mirallas, Navick, Obis,
  de~Solórzano, von Oy, Papaevangelou, Pérez, Picatoste, Ruz, Schiffer,
  Schmidt, Segui, and Vogel]{IAXO_FerrerRibas:2023}
E.~Ferrer-Ribas, K.~Altenmüller, B.~Biasuzzi, J.F. Castel, S.~Cebrián,
  T.~Dafni, K.~Desch, D.~Díez-Ibañez, J.~Galán, J.~Galindo, J.A. García,
  A.~Giganon, C.~Goblin, I.G. Irastorza, J.~Kaminski, G.~Luzón, C.~Margalejo,
  H.~Mirallas, X.F. Navick, L.~Obis, A.~Ortiz de~Solórzano, J.~von Oy,
  T.~Papaevangelou, O.~Pérez, E.~Picatoste, J.~Ruz, T.~Schiffer, S.~Schmidt,
  L.~Segui, and J.K. Vogel.
\newblock {U}ltra low background {M}icromegas detectors for {B}aby{IAXO} solar
  axion search.
\newblock \emph{Journal of Instrumentation}, 18\penalty0 (10):\penalty0 C10003,
  Oct 2023.
\newblock \doi{10.1088/1748-0221/18/10/C10003}.
\newblock URL \url{https://dx.doi.org/10.1088/1748-0221/18/10/C10003}.

\bibitem[Chen et~al.(2017)Chen, Fu, Galan, Giboni, Giuliani, Gu, Han, Ji, Lin,
  Liu, Ni, Kusano, Ren, Wang, Yang, Zhang, Zhang, Zhao, Sun, Hu, Jian, Li, Li,
  Liang, Zhang, Zhao, Zhou, Mao, Qiao, Wang, Yuan, Wang, Khan, Raper, Tang,
  Wang, Dong, Feng, Li, Liu, Liu, Wang, Zhu, Castel, Cebrián, Dafni, Garza,
  Irastorza, Iguaz, Luzón, Mirallas, Aune, Berthoumieux, Bedfer, Calvet,
  d'Hose, Delbart, Diakaki, Ferrer-Ribas, Ferrero, Kunne, Neyret,
  Papaevangelou, Sabatié, Vanderbroucke, Tan, Haxton, Mei, Kobdaj, and
  Yan]{PandaX-III:2017}
X.~Chen, C.~Fu, J.~Galan, K.~Giboni, F.~Giuliani, L.~Gu, K.~Han, X.~Ji, H.~Lin,
  J.~Liu, K.~Ni, H.~Kusano, X.~Ren, Sh. Wang, Y.~Yang, D.~Zhang, T.~Zhang,
  L.~Zhao, X.~Sun, S.~Hu, S.~Jian, X.~Li, X.~Li, H.~Liang, H.~Zhang, M.~Zhao,
  J.~Zhou, Y.~Mao, H.~Qiao, S.~Wang, Y.~Yuan, M.~Wang, A.~N. Khan, N.~Raper,
  J.~Tang, W.~Wang, J.~Dong, C.~Feng, C.~Li, J.~Liu, S.~Liu, X.~Wang, D.~Zhu,
  J.~F. Castel, S.~Cebrián, T.~Dafni, J.~G. Garza, I.~G. Irastorza, F.~J.
  Iguaz, G.~Luzón, H.~Mirallas, S.~Aune, E.~Berthoumieux, Y.~Bedfer,
  D.~Calvet, N.~d'Hose, A.~Delbart, M.~Diakaki, E.~Ferrer-Ribas, A.~Ferrero,
  F.~Kunne, D.~Neyret, T.~Papaevangelou, F.~Sabatié, M.~Vanderbroucke, A.~Tan,
  W.~Haxton, Y.~Mei, C.~Kobdaj, and Y.P. Yan.
\newblock {{P}anda{X-III}: {S}earching for neutrinoless double beta decay with
  high pressure $^{136}${X}e gas time projection chambers}.
\newblock \emph{Sci. China Phys. Mech. Astron.}, 60\penalty0 (6):\penalty0
  061011, 2017.
\newblock \doi{10.1007/s11433-017-9028-0}.
\newblock URL \url{https://doi.org/10.1007/s11433-017-9028-0}.

\bibitem[Iguaz et~al.(2016)Iguaz, Garza, Aznar, Castel, Cebrián, Dafni,
  García, Irastorza, Lagraba, Luzón, and Peiró]{TREX-DM:2016}
F.~J. Iguaz, J.~G. Garza, F.~Aznar, J.~F. Castel, S.~Cebrián, T.~Dafni, J.~A.
  García, I.~G. Irastorza, A.~Lagraba, G.~Luzón, and A.~Peiró.
\newblock {TREX-DM}: a low-background {M}icromegas-based {TPC} for low-mass
  {WIMP} detection.
\newblock \emph{Eur. Phys. J. C}, 76\penalty0 (10):\penalty0 529, 2016.
\newblock \doi{10.1140/epjc/s10052-016-4372-6}.
\newblock URL \url{https://doi.org/10.1140/epjc/s10052-016-4372-6}.

\bibitem[Castel et~al.(2020)Castel, Cebrián, Dafni, Galán, Irastorza, Luzón,
  Margalejo, Mirallas, de~Solórzano, Peiró, and Ruiz-Chóliz]{TREX-DM:2020}
J.~Castel, S.~Cebrián, T.~Dafni, J.~Galán, I.G. Irastorza, G.~Luzón,
  C.~Margalejo, H.~Mirallas, A.~Ortiz de~Solórzano, A.~Peiró, and
  E.~Ruiz-Chóliz.
\newblock {T}he {TREX-DM} experiment at the {C}anfranc {U}nderground
  {L}aboratory.
\newblock \emph{Journal of Physics: Conference Series}, 1468\penalty0
  (1):\penalty0 012063, Feb 2020.
\newblock \doi{10.1088/1742-6596/1468/1/012063}.
\newblock URL \url{https://dx.doi.org/10.1088/1742-6596/1468/1/012063}.

\bibitem[Luzón et~al.(2024)Luzón, Dafni, Altenmueller, Antolín, Calvet,
  Candón, Cebrián, Castel, Cogollos, Díez~Ibáñez, Ferrer-Ribas, Galán,
  Antonio~García, Gómez, Gu, Ezquerro, García~Irastorza, Margalejo,
  Mirallas, Porrón, Quintana, Obis, Ortiz~de Solorzano, Papaevangelou, Pérez,
  Picatoste, Jiménez~Puyuelo, Ruiz-Chóliz, Ruz, and
  Vogel]{MicromegasPerspectives:2024}
G.~Luzón, T.~Dafni, K.~Altenmueller, I.~Antolín, D.~Calvet, F.~Candón,
  S.~Cebrián, J.~Castel, C.~Cogollos, D.~Díez~Ibáñez, E.~Ferrer-Ribas,
  J.~Galán, J.~Antonio~García, H.~Gómez, Y.~Gu, Á. Ezquerro,
  I.~García~Irastorza, C.~Margalejo, H.~Mirallas, J.~Porrón, A.~Quintana,
  L.~Obis, A.~Ortiz~de Solorzano, T.~Papaevangelou, Ó. Pérez, E.~Picatoste,
  M.~Jiménez~Puyuelo, E.~Ruiz-Chóliz, J.~Ruz, and J.~Vogel.
\newblock {U}sing {M}icromegas detectors for direct dark matter searches:
  challenges and perspectives.
\newblock \emph{Journal of Advanced Instrumentation in Science}, 2024\penalty0
  (1), Nov 2024.
\newblock \doi{10.31526/jais.2024.549}.
\newblock URL
  \url{https://jais.andromedapublisher.org/index.php/JAIS/article/view/549}.

\bibitem[Lewin and Smith(1996)]{Lewin:1996}
J.D. Lewin and P.F. Smith.
\newblock {R}eview of mathematics, numerical factors, and corrections for dark
  matter experiments based on elastic nuclear recoil.
\newblock \emph{Astroparticle Physics}, 6\penalty0 (1):\penalty0 87--112, 1996.
\newblock ISSN 0927-6505.
\newblock \doi{10.1016/S0927-6505(96)00047-3}.
\newblock URL
  \url{https://www.sciencedirect.com/science/article/pii/S0927650596000473}.

\bibitem[Derré et~al.(2000)Derré, Giomataris, Rebourgeard, Zaccone, Perroud,
  and Charpak]{Derre:2000}
J.~Derré, Y.~Giomataris, Ph. Rebourgeard, H.~Zaccone, J.P. Perroud, and
  G.~Charpak.
\newblock {F}ast signals and single electron detection with a {MICROMEGAS}
  photodetector.
\newblock \emph{Nuclear Instruments and Methods in Physics Research Section A:
  Accelerators, Spectrometers, Detectors and Associated Equipment},
  449\penalty0 (1):\penalty0 314--321, 2000.
\newblock ISSN 0168-9002.
\newblock \doi{10.1016/S0168-9002(99)01452-7}.
\newblock URL
  \url{https://www.sciencedirect.com/science/article/pii/S0168900299014527}.

\bibitem[Kane et~al.(2002)Kane, May, Miyamoto, Shipsey, Andriamonje, Delbart,
  Derre, Giomataris, and Jeanneau]{GEM+MM:2002}
S.~Kane, J.~May, J.~Miyamoto, I.~Shipsey, S.~Andriamonje, A.~Delbart, J.~Derre,
  I.~Giomataris, and F.~Jeanneau.
\newblock {A} study of {M}icromegas with preamplification with a single {GEM}.
\newblock In M.~Barone, E.~Borchi, J.~Huston, C.~Leroy, P.~G. Rancoita,
  P.~Riboni, and R.~Ruchti, editors, \emph{Advanced Technology - Particle
  Physics}, pages 694--703, Nov 2002.
\newblock \doi{10.1142/9789812776464_0098}.
\newblock URL \url{https://dx.doi.org/10.1142/9789812776464_0098}.

\bibitem[Neyret et~al.(2009)Neyret, Anfreville, Bedfer, Burtin, d'Hose,
  Giganon, Ketzer, Konorov, Kunne, Magnon, Marchand, Paul, Platchkov, and
  Vandenbroucke]{Neyret:2009}
D.~Neyret, M.~Anfreville, Y.~Bedfer, E.~Burtin, N.~d'Hose, A.~Giganon,
  B.~Ketzer, I.~Konorov, F.~Kunne, A.~Magnon, C.~Marchand, B.~Paul,
  S.~Platchkov, and M.~Vandenbroucke.
\newblock {N}ew pixelized {M}icromegas detector for the {COMPASS} experiment.
\newblock \emph{Journal of Instrumentation}, 4\penalty0 (12):\penalty0 P12004,
  dec 2009.
\newblock \doi{10.1088/1748-0221/4/12/P12004}.
\newblock URL \url{https://dx.doi.org/10.1088/1748-0221/4/12/P12004}.

\bibitem[Neyret et~al.(2024)Neyret, Abbon, Anfreville, Andrieux, Bedfer,
  Durand, Herlant, d’Hose, Kunne, Platchkov, Thibaud, Usseglio, and
  Vandenbroucke]{Neyret:2024}
D.~Neyret, P.~Abbon, M.~Anfreville, V.~Andrieux, Y.~Bedfer, D.~Durand,
  S.~Herlant, N.~d’Hose, F.~Kunne, S.~Platchkov, F.~Thibaud, M.~Usseglio, and
  M.~Vandenbroucke.
\newblock {A}ging effects in the {COMPASS} hybrid {GEM}-{M}icromegas pixelized
  detectors.
\newblock \emph{Nuclear Instruments and Methods in Physics Research Section A:
  Accelerators, Spectrometers, Detectors and Associated Equipment},
  1065:\penalty0 169511, 2024.
\newblock ISSN 0168-9002.
\newblock \doi{https://doi.org/10.1016/j.nima.2024.169511}.
\newblock URL
  \url{https://www.sciencedirect.com/science/article/pii/S0168900224004376}.

\bibitem[Anvar et~al.(2011)Anvar, Baron, Blank, Chavas, Delagnes, Druillole,
  Hellmuth, Nalpas, Pedroza, Pibernat, Pollacco, Rebii, and Usher]{AGET:2011}
S.~Anvar, P.~Baron, B.~Blank, J.~Chavas, E.~Delagnes, F.~Druillole,
  P.~Hellmuth, L.~Nalpas, J.L. Pedroza, J.~Pibernat, E.~Pollacco, A.~Rebii, and
  N.~Usher.
\newblock {AGET}, the {GET} front-end {ASIC}, for the readout of the {T}ime
  {P}rojection {C}hambers used in nuclear physic experiments.
\newblock In \emph{2011 IEEE Nuclear Science Symposium Conference Record},
  pages 745--749, 2011.
\newblock \doi{10.1109/NSSMIC.2011.6154095}.
\newblock URL \url{https://dx.doi.org/10.1109/NSSMIC.2011.6154095}.

\bibitem[Calvet(2014)]{FEMINOS:2014}
D.~Calvet.
\newblock {A} {V}ersatile {R}eadout {S}ystem for {S}mall to {M}edium {S}cale
  {G}aseous and {S}ilicon {D}etectors.
\newblock \emph{IEEE Transactions on Nuclear Science}, 61\penalty0
  (1):\penalty0 675--682, 2014.
\newblock \doi{10.1109/TNS.2014.2299312}.
\newblock URL \url{https://dx.doi.org/10.1109/TNS.2014.2299312}.

\bibitem[Altenmüller et~al.(2022)Altenmüller, Cebrián, Dafni,
  Díez-Ibáñez, Galán, Galindo, García, Irastorza, Luzón, Margalejo,
  Mirallas, Obis, Pérez, Han, Ni, Bedfer, Biasuzzi, Ferrer-Ribas, Neyret,
  Papaevangelou, Cogollos, and Picatoste]{REST:2022}
K.~Altenmüller, S.~Cebrián, T.~Dafni, D.~Díez-Ibáñez, J.~Galán,
  J.~Galindo, J.~Antonio García, I.~G. Irastorza, G.~Luzón, C.~Margalejo,
  H.~Mirallas, L.~Obis, O.~Pérez, K.~Han, K.~Ni, Y.~Bedfer, B.~Biasuzzi,
  E.~Ferrer-Ribas, D.~Neyret, T.~Papaevangelou, C.~Cogollos, and E.~Picatoste.
\newblock {REST}-for-{P}hysics, a {ROOT}-based framework for event oriented
  data analysis and combined {M}onte {C}arlo response.
\newblock \emph{Computer Physics Communications}, 273:\penalty0 108281, 2022.
\newblock ISSN 0010-4655.
\newblock \doi{10.1016/j.cpc.2021.108281}.
\newblock URL
  \url{https://www.sciencedirect.com/science/article/pii/S0010465521003933}.

\bibitem[Abeln et~al.(2021)Abeln, Altenmüller, Cuendis, Armengaud, Attié,
  Aune, Basso, Bergé, Biasuzzi, Sousa, Brun, Bykovskiy, Calvet, Carmona,
  Castel, Cebrián, Chernov, Christensen, Civitani, Cogollos, Dafní, Derbin,
  Desch, Díez, Dinter, Döbrich, Drachnev, Dudarev, Dumoulin, Ferreira,
  Ferrer-Ribas, Fleck, Galán, Gascón, Gastaldo, Giannotti, Giomataris,
  Giuliani, Gninenko, Golm, Golubev, Hagge, Hahn, Hailey, Hengstler, Henriksen,
  Houdy, Iglesias-Marzoa, Iguaz, Irastorza, Iñiguez, Jakovcic, Kaminski,
  Kanoute, Karstensen, Kravchuk, Lakic, Lasserre, Laurent, Limousin, Lindner,
  Loidl, Lomskaya, López-Alegre, Lubsandorzhiev, Ludwig, Luzón, Malbrunot,
  Margalejo, Marin-Franch, Marnieros, Marutzky, Mauricio, Menesguen, Mentink,
  Mertens, Mescia, Miralda-Escudé, Mirallas, Mols, Muratova, Navick, Nones,
  Notari, Nozik, Obis, Oriol, Orsini, de~Solórzano, Oster, Silva, Pantuev,
  Papaevangelou, Pareschi, Perez, Pérez, Picatoste, Pivovaroff, Poda, Redondo,
  Ringwald, Rodrigues, Rueda-Teruel, Rueda-Teruel, Ruiz-Choliz, Ruz, Saemann,
  Salvado, Schiffer, Schmidt, Schneekloth, Schott, Segui, Tavecchio, ten Kate,
  Tkachev, Troitsky, Unger, Unzhakov, Ushakov, Vogel, Voronin, Weltman,
  Werthenbach, Wuensch, and Yanes-Díaz]{babyIAXO:2020}
A.~Abeln, K.~Altenmüller, S.~Arguedas Cuendis, E.~Armengaud, D.~Attié,
  S.~Aune, S.~Basso, L.~Bergé, B.~Biasuzzi, P.~T. C. Borges~De Sousa, P.~Brun,
  N.~Bykovskiy, D.~Calvet, J.~M. Carmona, J.~F. Castel, S.~Cebrián,
  V.~Chernov, F.~E. Christensen, M.~M. Civitani, C.~Cogollos, T.~Dafní,
  A.~Derbin, K.~Desch, D.~Díez, M.~Dinter, B.~Döbrich, I.~Drachnev,
  A.~Dudarev, L.~Dumoulin, D.~D.~M. Ferreira, E.~Ferrer-Ribas, I.~Fleck,
  J.~Galán, D.~Gascón, L.~Gastaldo, M.~Giannotti, Y.~Giomataris, A.~Giuliani,
  S.~Gninenko, J.~Golm, N.~Golubev, L.~Hagge, J.~Hahn, C.~J. Hailey,
  D.~Hengstler, P.~L. Henriksen, T.~Houdy, R.~Iglesias-Marzoa, F.~J. Iguaz,
  I.~G. Irastorza, C.~Iñiguez, K.~Jakovcic, J.~Kaminski, B.~Kanoute,
  S.~Karstensen, L.~Kravchuk, B.~Lakic, T.~Lasserre, P.~Laurent, O.~Limousin,
  A.~Lindner, M.~Loidl, I.~Lomskaya, G.~López-Alegre, B.~Lubsandorzhiev,
  K.~Ludwig, G.~Luzón, C.~Malbrunot, C.~Margalejo, A.~Marin-Franch,
  S.~Marnieros, F.~Marutzky, J.~Mauricio, Y.~Menesguen, M.~Mentink, S.~Mertens,
  F.~Mescia, J.~Miralda-Escudé, H.~Mirallas, J.~P. Mols, V.~Muratova, X.~F.
  Navick, C.~Nones, A.~Notari, A.~Nozik, L.~Obis, C.~Oriol, F.~Orsini, A.~Ortiz
  de~Solórzano, S.~Oster, H.~P. Pais~Da Silva, V.~Pantuev, T.~Papaevangelou,
  G.~Pareschi, K.~Perez, O.~Pérez, E.~Picatoste, M.~J. Pivovaroff, D.~V. Poda,
  J.~Redondo, A.~Ringwald, M.~Rodrigues, F.~Rueda-Teruel, S.~Rueda-Teruel,
  E.~Ruiz-Choliz, J.~Ruz, E.~O. Saemann, J.~Salvado, T.~Schiffer, S.~Schmidt,
  U.~Schneekloth, M.~Schott, L.~Segui, F.~Tavecchio, H.~H.~J. ten Kate,
  I.~Tkachev, S.~Troitsky, D.~Unger, E.~Unzhakov, N.~Ushakov, J.~K. Vogel,
  D.~Voronin, A.~Weltman, U.~Werthenbach, W.~Wuensch, and A.~Yanes-Díaz.
\newblock {{C}onceptual design of {B}aby{IAXO}, the intermediate stage towards
  the {I}nternational {A}xion {O}bservatory}.
\newblock \emph{JHEP}, 05:\penalty0 137, 2021.
\newblock \doi{10.1007/JHEP05(2021)137}.
\newblock URL \url{https://dx.doi.org/10.1007/JHEP05(2021)137}.

\bibitem[Castel et~al.(2019)Castel, Cebrián, Coarasa, Dafni, Galan, Iguaz,
  Irastorza, Luzón, Mirallas, Ortiz~de Solórzano, and
  Ruiz-Chóliz]{TREXDM_Bckg_Assessment:2018}
J.~F. Castel, S.~Cebrián, I.~Coarasa, T.~Dafni, J.~Galan, F.~Iguaz,
  I.~Irastorza, G.~Luzón, H.~Mirallas, A.~Ortiz~de Solórzano, and
  E.~Ruiz-Chóliz.
\newblock {B}ackground assessment for the {TREX} dark matter experiment.
\newblock \emph{Eur. Phys. J. C}, 79\penalty0 (9):\penalty0 782, 2019.
\newblock \doi{10.1140/epjc/s10052-019-7282-6}.
\newblock URL \url{https://dx.doi.org/10.1140/epjc/s10052-019-7282-6}.

\bibitem[Iguaz et~al.(2022)Iguaz, Dafni, Canellas, Castel, Cebrián, Garza,
  Irastorza, Luzón, Mirallas, and Ruiz-Chóliz]{Iguaz:2022}
F.J. Iguaz, T.~Dafni, C.~Canellas, J.F. Castel, S.~Cebrián, J.G. Garza, I.G.
  Irastorza, G.~Luzón, H.~Mirallas, and E.~Ruiz-Chóliz.
\newblock {M}icrobulk {M}icromegas in non-flammable mixtures of argon and neon
  at high pressure.
\newblock \emph{Journal of Instrumentation}, 17\penalty0 (07):\penalty0 P07032,
  Jul 2022.
\newblock \doi{10.1088/1748-0221/17/07/P07032}.
\newblock URL \url{https://dx.doi.org/10.1088/1748-0221/17/07/P07032}.

\bibitem[Berger et~al.(2010)Berger, Hubbell, Seltzer, Chang, Coursey, Sukumar,
  Zucker, and Olsen]{NIST_XCOM:2010}
M.J. Berger, J.H. Hubbell, S.M. Seltzer, J.~Chang, J.S. Coursey, R.~Sukumar,
  D.S. Zucker, and K.~Olsen.
\newblock {XCOM}: {P}hoton {C}ross {S}ection {D}atabase (version 1.5).
\newblock \url{http://physics.nist.gov/xcom}, 2010.

\bibitem[Sauli et~al.(2002)Sauli, Kappler, and Ropelewski]{Sauli:2002}
F.~Sauli, S.~Kappler, and L.~Ropelewski.
\newblock {E}lectron collection and ion feedback in {GEM}-based detectors.
\newblock In \emph{2002 IEEE Nuclear Science Symposium Conference Record},
  volume~1, pages 278--282, 2002.
\newblock \doi{10.1109/NSSMIC.2002.1239316}.
\newblock URL \url{https://dx.doi.org/10.1109/NSSMIC.2002.1239316}.

\bibitem[Bencivenni et~al.(2003)Bencivenni, Bonivento, Cardini, Deplano,
  De~Simone, Murtas, Pinci, Poli-Lener, and Raspino]{GEM:2003}
G.~Bencivenni, W.~Bonivento, A.~Cardini, C.~Deplano, P.~De~Simone, F.~Murtas,
  D.~Pinci, M.~Poli-Lener, and D.~Raspino.
\newblock {M}easurement of {GEM} parameters with {X}-rays.
\newblock \emph{IEEE Transactions on Nuclear Science}, 50\penalty0
  (5):\penalty0 1297--1302, 2003.
\newblock \doi{10.1109/TNS.2003.818234}.
\newblock URL \url{https://dx.doi.org/10.1109/TNS.2003.818234}.

\bibitem[Peisert and Sauli(1984)]{diffusion:1984}
A.~Peisert and F.~Sauli.
\newblock \emph{{D}rift and diffusion of electrons in gases: a compilation
  (with an introduction to the use of computing programs)}.
\newblock CERN Yellow Reports: Monographs. CERN, Geneva, 1984.
\newblock \doi{10.5170/CERN-1984-008}.
\newblock URL \url{https://cds.cern.ch/record/154069}.

\bibitem[Atti\'e et~al.(2013)Atti\'e, Chaus, Colas, Ribas, Gal\'an, Giomataris,
  Gongadze, Iguaz, Oliveira, Papaevangelou, and Peyaud]{Attie:2013}
D.~Atti\'e, A.~Chaus, P.~Colas, E.~Ferrer Ribas, J.~Gal\'an, I.~Giomataris,
  A.~Gongadze, F.~J. Iguaz, R.~De Oliveira, T.~Papaevangelou, and A.~Peyaud.
\newblock {A} {P}iggyback resistive {M}icromegas.
\newblock \emph{Journal of Instrumentation}, 8\penalty0 (05):\penalty0 P05019,
  may 2013.
\newblock \doi{10.1088/1748-0221/8/05/P05019}.
\newblock URL \url{https://dx.doi.org/10.1088/1748-0221/8/05/P05019}.

\bibitem[Agnese et~al.(2019)Agnese, Aralis, Aramaki, Arnquist, Azadbakht,
  Baker, Banik, Barker, Bauer, Binder, Bowles, Brink, Bunker, Cabrera, Calkins,
  Cameron, Cartaro, Cerde\~no, Chang, Cooley, Cornell, Cushman, De~Brienne,
  Doughty, Fascione, Figueroa-Feliciano, Fink, Fritts, Gerbier, Germond,
  Ghaith, Golwala, Harris, Herbert, Hong, Hoppe, Hsu, Huber, Iyer, Jardin,
  Jastram, Jena, Kelsey, Kennedy, Kubik, Kurinsky, Lawrence, Loer,
  Lopez~Asamar, Lukens, MacDonell, Mahapatra, Mandic, Mast, Miller,
  Mirabolfathi, Mohanty, Morales~Mendoza, Nelson, Neog, Orrell, Oser, Page,
  Partridge, Pepin, Ponce, Poudel, Pyle, Qiu, Rau, Reisetter, Ren, Reynolds,
  Roberts, Robinson, Rogers, Saab, Sadoulet, Sander, Scarff, Schnee, Scorza,
  Senapati, Serfass, Speller, Stanford, Stein, Street, Tanaka, Toback,
  Underwood, Villano, von Krosigk, Watkins, Wilson, Wilson, Winchell, Wright,
  Yellin, Young, Zhang, and Zhao]{CDMSLite:2018}
R.~Agnese, T.~Aralis, T.~Aramaki, I.~J. Arnquist, E.~Azadbakht, W.~Baker,
  S.~Banik, D.~Barker, D.~A. Bauer, T.~Binder, M.~A. Bowles, P.~L. Brink,
  R.~Bunker, B.~Cabrera, R.~Calkins, R.~A. Cameron, C.~Cartaro, D.~G.
  Cerde\~no, Y.-Y. Chang, J.~Cooley, B.~Cornell, P.~Cushman, F.~De~Brienne,
  T.~Doughty, E.~Fascione, E.~Figueroa-Feliciano, C.~W. Fink, M.~Fritts,
  G.~Gerbier, R.~Germond, M.~Ghaith, S.~R. Golwala, H.~R. Harris, N.~Herbert,
  Z.~Hong, E.~W. Hoppe, L.~Hsu, M.~E. Huber, V.~Iyer, D.~Jardin, A.~Jastram,
  C.~Jena, M.~H. Kelsey, A.~Kennedy, A.~Kubik, N.~A. Kurinsky, R.~E. Lawrence,
  B.~Loer, E.~Lopez~Asamar, P.~Lukens, D.~MacDonell, R.~Mahapatra, V.~Mandic,
  N.~Mast, E.~Miller, N.~Mirabolfathi, B.~Mohanty, J.~D. Morales~Mendoza,
  J.~Nelson, H.~Neog, J.~L. Orrell, S.~M. Oser, W.~A. Page, R.~Partridge,
  M.~Pepin, F.~Ponce, S.~Poudel, M.~Pyle, H.~Qiu, W.~Rau, A.~Reisetter, R.~Ren,
  T.~Reynolds, A.~Roberts, A.~E. Robinson, H.~E. Rogers, T.~Saab, B.~Sadoulet,
  J.~Sander, A.~Scarff, R.~W. Schnee, S.~Scorza, K.~Senapati, B.~Serfass,
  D.~Speller, C.~Stanford, M.~Stein, J.~Street, H.~A. Tanaka, D.~Toback,
  R.~Underwood, A.~N. Villano, B.~von Krosigk, S.~L. Watkins, J.~S. Wilson,
  M.~J. Wilson, J.~Winchell, D.~H. Wright, S.~Yellin, B.~A. Young, X.~Zhang,
  and X.~Zhao.
\newblock {S}earch for low-mass dark matter with {CDMS}lite using a profile
  likelihood fit.
\newblock \emph{Phys. Rev. D}, 99:\penalty0 062001, Mar 2019.
\newblock \doi{10.1103/PhysRevD.99.062001}.
\newblock URL \url{https://link.aps.org/doi/10.1103/PhysRevD.99.062001}.

\bibitem[Aprile et~al.(2018)Aprile, Aalbers, Agostini, Alfonsi, Althueser,
  Amaro, Anthony, Arneodo, Baudis, Bauermeister, Benabderrahmane, Berger,
  Breur, Brown, Brown, Brown, Bruenner, Bruno, Budnik, Capelli, Cardoso,
  Cichon, Coderre, Colijn, Conrad, Cussonneau, Decowski, de~Perio, Di~Gangi,
  Di~Giovanni, Diglio, Elykov, Eurin, Fei, Ferella, Fieguth, Fulgione,
  Gallo~Rosso, Galloway, Gao, Garbini, Geis, Grandi, Greene, Qiu, Hasterok,
  Hogenbirk, Howlett, Itay, Joerg, Kaminsky, Kazama, Kish, Koltman, Landsman,
  Lang, Levinson, Lin, Lindemann, Lindner, Lombardi, Lopes, Mahlstedt,
  Manfredini, Marrod\'an~Undagoitia, Masbou, Masson, Messina, Micheneau,
  Miller, Molinario, Mor\aa{}, Murra, Naganoma, Ni, Oberlack, Pelssers,
  Piastra, Pienaar, Pizzella, Plante, Podviianiuk, Priel,
  Ram\'{\i}rez~Garc\'{\i}a, Rauch, Reichard, Reuter, Riedel, Rizzo, Rocchetti,
  Rupp, dos Santos, Sartorelli, Scheibelhut, Schindler, Schreiner, Schulte,
  Schumann, Scotto~Lavina, Selvi, Shagin, Shockley, Silva, Simgen, Thers,
  Toschi, Trinchero, Tunnell, Upole, Vargas, Wack, Wang, Wang, Wei, Weinheimer,
  Wittweg, Wulf, Ye, Zhang, and Zhu]{XENON1T:2018}
E.~Aprile, J.~Aalbers, F.~Agostini, M.~Alfonsi, L.~Althueser, F.~D. Amaro,
  M.~Anthony, F.~Arneodo, L.~Baudis, B.~Bauermeister, M.~L. Benabderrahmane,
  T.~Berger, P.~A. Breur, A.~Brown, A.~Brown, E.~Brown, S.~Bruenner, G.~Bruno,
  R.~Budnik, C.~Capelli, J.~M.~R. Cardoso, D.~Cichon, D.~Coderre, A.~P. Colijn,
  J.~Conrad, J.~P. Cussonneau, M.~P. Decowski, P.~de~Perio, P.~Di~Gangi,
  A.~Di~Giovanni, S.~Diglio, A.~Elykov, G.~Eurin, J.~Fei, A.~D. Ferella,
  A.~Fieguth, W.~Fulgione, A.~Gallo~Rosso, M.~Galloway, F.~Gao, M.~Garbini,
  C.~Geis, L.~Grandi, Z.~Greene, H.~Qiu, C.~Hasterok, E.~Hogenbirk, J.~Howlett,
  R.~Itay, F.~Joerg, B.~Kaminsky, S.~Kazama, A.~Kish, G.~Koltman, H.~Landsman,
  R.~F. Lang, L.~Levinson, Q.~Lin, S.~Lindemann, M.~Lindner, F.~Lombardi,
  J.~A.~M. Lopes, J.~Mahlstedt, A.~Manfredini, T.~Marrod\'an~Undagoitia,
  J.~Masbou, D.~Masson, M.~Messina, K.~Micheneau, K.~Miller, A.~Molinario,
  K.~Mor\aa{}, M.~Murra, J.~Naganoma, K.~Ni, U.~Oberlack, B.~Pelssers,
  F.~Piastra, J.~Pienaar, V.~Pizzella, G.~Plante, R.~Podviianiuk, N.~Priel,
  D.~Ram\'{\i}rez~Garc\'{\i}a, L.~Rauch, S.~Reichard, C.~Reuter, B.~Riedel,
  A.~Rizzo, A.~Rocchetti, N.~Rupp, J.~M.~F. dos Santos, G.~Sartorelli,
  M.~Scheibelhut, S.~Schindler, J.~Schreiner, D.~Schulte, M.~Schumann,
  L.~Scotto~Lavina, M.~Selvi, P.~Shagin, E.~Shockley, M.~Silva, H.~Simgen,
  D.~Thers, F.~Toschi, G.~Trinchero, C.~Tunnell, N.~Upole, M.~Vargas, O.~Wack,
  H.~Wang, Z.~Wang, Y.~Wei, C.~Weinheimer, C.~Wittweg, J.~Wulf, J.~Ye,
  Y.~Zhang, and T.~Zhu.
\newblock {D}ark {M}atter {S}earch {R}esults from a {O}ne {T}on-{Y}ear
  {E}xposure of {XENON1T}.
\newblock \emph{Phys. Rev. Lett.}, 121:\penalty0 111302, Sep 2018.
\newblock \doi{10.1103/PhysRevLett.121.111302}.
\newblock URL \url{https://link.aps.org/doi/10.1103/PhysRevLett.121.111302}.

\bibitem[Abdelhameed et~al.(2019)Abdelhameed, Angloher, Bauer, Bento, Bertoldo,
  Bucci, Canonica, D'Addabbo, Defay, Di~Lorenzo, Erb, Feilitzsch, Fichtinger,
  Ferreiro~Iachellini, Fuss, Gorla, Hauff, Jochum, Kinast, Kluck, Kraus,
  Langenk\"amper, Mancuso, Mokina, Mondragon, M\"unster, Olmi, Ortmann,
  Pagliarone, Pattavina, Petricca, Potzel, Pr\"obst, Reindl, Rothe,
  Sch\"affner, Schieck, Schipperges, Schmiedmayer, Sch\"onert, Schwertner,
  Stahlberg, Stodolsky, Strandhagen, Strauss, T\"urko\ifmmode~\check{g}\else
  \v{g}\fi{}lu, Usherov, Willers, and Zema]{CREST-III:2019}
A.~H. Abdelhameed, G.~Angloher, P.~Bauer, A.~Bento, E.~Bertoldo, C.~Bucci,
  L.~Canonica, A.~D'Addabbo, X.~Defay, S.~Di~Lorenzo, A.~Erb, F.~v. Feilitzsch,
  S.~Fichtinger, N.~Ferreiro~Iachellini, A.~Fuss, P.~Gorla, D.~Hauff,
  J.~Jochum, A.~Kinast, H.~Kluck, H.~Kraus, A.~Langenk\"amper, M.~Mancuso,
  V.~Mokina, E.~Mondragon, A.~M\"unster, M.~Olmi, T.~Ortmann, C.~Pagliarone,
  L.~Pattavina, F.~Petricca, W.~Potzel, F.~Pr\"obst, F.~Reindl, J.~Rothe,
  K.~Sch\"affner, J.~Schieck, V.~Schipperges, D.~Schmiedmayer, S.~Sch\"onert,
  C.~Schwertner, M.~Stahlberg, L.~Stodolsky, C.~Strandhagen, R.~Strauss,
  C.~T\"urko\ifmmode~\check{g}\else \v{g}\fi{}lu, I.~Usherov, M.~Willers, and
  V.~Zema.
\newblock {F}irst results from the {CRESST-III} low-mass dark matter program.
\newblock \emph{Phys. Rev. D}, 100:\penalty0 102002, Nov 2019.
\newblock \doi{10.1103/PhysRevD.100.102002}.
\newblock URL \url{https://link.aps.org/doi/10.1103/PhysRevD.100.102002}.

\bibitem[Aguilar-Arevalo et~al.(2020)Aguilar-Arevalo, Amidei, Baxter, Cancelo,
  Vergara, Chavarria, D'Olivo, Estrada, Favela-Perez, Ga\"{\i}or, Guardincerri,
  Hoppe, Hossbach, Kilminster, Lawson, Lee, Letessier-Selvon, Matalon, Mitra,
  Overman, Piers, Privitera, Ramanathan, Da~Rocha, Sarkis, Settimo, Smida,
  Thomas, Tiffenberg, Traina, Vilar, and Virto]{DAMIC:2020}
A.~Aguilar-Arevalo, D.~Amidei, D.~Baxter, G.~Cancelo, B.~A.~Cervantes Vergara,
  A.~E. Chavarria, J.~C. D'Olivo, J.~Estrada, F.~Favela-Perez, R.~Ga\"{\i}or,
  Y.~Guardincerri, E.~W. Hoppe, T.~W. Hossbach, B.~Kilminster, I.~Lawson, S.~J.
  Lee, A.~Letessier-Selvon, A.~Matalon, P.~Mitra, C.~T. Overman, A.~Piers,
  P.~Privitera, K.~Ramanathan, J.~Da~Rocha, Y.~Sarkis, M.~Settimo, R.~Smida,
  R.~Thomas, J.~Tiffenberg, M.~Traina, R.~Vilar, and A.~L. Virto.
\newblock {R}esults on {L}ow-{M}ass {W}eakly {I}nteracting {M}assive
  {P}articles from an 11 kg d {T}arget {E}xposure of {DAMIC} at {SNOLAB}.
\newblock \emph{Phys. Rev. Lett.}, 125:\penalty0 241803, Dec 2020.
\newblock \doi{10.1103/PhysRevLett.125.241803}.
\newblock URL \url{https://link.aps.org/doi/10.1103/PhysRevLett.125.241803}.

\bibitem[Aprile et~al.(2021)Aprile, Aalbers, Agostini, Ahmed~Maouloud, Alfonsi,
  Althueser, Amaro, Andaloro, Antochi, Angelino, Angevaare, Arneodo, Baudis,
  Bauermeister, Bellagamba, Benabderrahmane, Brown, Brown, Bruenner, Bruno,
  Budnik, Capelli, Cardoso, Cichon, Cimmino, Clark, Coderre, Colijn, Conrad,
  Cuenca, Cussonneau, Decowski, Depoian, Di~Gangi, Di~Giovanni, Di~Stefano,
  Diglio, Elykov, Ferella, Fulgione, Gaemers, Gaior, Galloway, Gao, Grandi,
  Hils, Hiraide, Hoetzsch, Howlett, Iacovacci, Itow, Joerg, Kato, Kazama,
  Kobayashi, Koltman, Kopec, Landsman, Lang, Levinson, Liang, Lindemann,
  Lindner, Lombardi, Long, Lopes, Ma, Macolino, Mahlstedt, Mancuso, Manenti,
  Manfredini, Marignetti, Marrod\'an~Undagoitia, Martens, Masbou, Masson,
  Mastroianni, Messina, Miuchi, Mizukoshi, Molinario, Mor\aa{}, Moriyama,
  Mosbacher, Murra, Naganoma, Ni, Oberlack, Odgers, Palacio, Pelssers, Peres,
  Pierre, Pienaar, Pizzella, Plante, Qi, Qin, Ram\'{\i}rez~Garc\'{\i}a,
  Reichard, Rocchetti, Rupp, dos Santos, Sartorelli, Schreiner, Schulte,
  Schulze~Ei\ss{}ing, Schumann, Scotto~Lavina, Selvi, Semeria, Shagin,
  Shockley, Silva, Simgen, Takeda, Therreau, Thers, Toschi, Trinchero, Tunnell,
  Valerius, Vargas, Volta, Wei, Weinheimer, Weiss, Wenz, Wittweg, Wolf, Xu,
  Yamashita, Ye, Zavattini, Zhang, Zhu, and Zopounidis]{XENON1T:2021}
E.~Aprile, J.~Aalbers, F.~Agostini, S.~Ahmed~Maouloud, M.~Alfonsi,
  L.~Althueser, F.~D. Amaro, S.~Andaloro, V.~C. Antochi, E.~Angelino, J.~R.
  Angevaare, F.~Arneodo, L.~Baudis, B.~Bauermeister, L.~Bellagamba, M.~L.
  Benabderrahmane, A.~Brown, E.~Brown, S.~Bruenner, G.~Bruno, R.~Budnik,
  C.~Capelli, J.~M.~R. Cardoso, D.~Cichon, B.~Cimmino, M.~Clark, D.~Coderre,
  A.~P. Colijn, J.~Conrad, J.~Cuenca, J.~P. Cussonneau, M.~P. Decowski,
  A.~Depoian, P.~Di~Gangi, A.~Di~Giovanni, R.~Di~Stefano, S.~Diglio, A.~Elykov,
  A.~D. Ferella, W.~Fulgione, P.~Gaemers, R.~Gaior, M.~Galloway, F.~Gao,
  L.~Grandi, C.~Hils, K.~Hiraide, L.~Hoetzsch, J.~Howlett, M.~Iacovacci,
  Y.~Itow, F.~Joerg, N.~Kato, S.~Kazama, M.~Kobayashi, G.~Koltman, A.~Kopec,
  H.~Landsman, R.~F. Lang, L.~Levinson, S.~Liang, S.~Lindemann, M.~Lindner,
  F.~Lombardi, J.~Long, J.~A.~M. Lopes, Y.~Ma, C.~Macolino, J.~Mahlstedt,
  A.~Mancuso, L.~Manenti, A.~Manfredini, F.~Marignetti,
  T.~Marrod\'an~Undagoitia, K.~Martens, J.~Masbou, D.~Masson, S.~Mastroianni,
  M.~Messina, K.~Miuchi, K.~Mizukoshi, A.~Molinario, K.~Mor\aa{}, S.~Moriyama,
  Y.~Mosbacher, M.~Murra, J.~Naganoma, K.~Ni, U.~Oberlack, K.~Odgers,
  J.~Palacio, B.~Pelssers, R.~Peres, M.~Pierre, J.~Pienaar, V.~Pizzella,
  G.~Plante, J.~Qi, J.~Qin, D.~Ram\'{\i}rez~Garc\'{\i}a, S.~Reichard,
  A.~Rocchetti, N.~Rupp, J.~M.~F. dos Santos, G.~Sartorelli, J.~Schreiner,
  D.~Schulte, H.~Schulze~Ei\ss{}ing, M.~Schumann, L.~Scotto~Lavina, M.~Selvi,
  F.~Semeria, P.~Shagin, E.~Shockley, M.~Silva, H.~Simgen, A.~Takeda,
  C.~Therreau, D.~Thers, F.~Toschi, G.~Trinchero, C.~Tunnell, K.~Valerius,
  M.~Vargas, G.~Volta, Y.~Wei, C.~Weinheimer, M.~Weiss, D.~Wenz, C.~Wittweg,
  T.~Wolf, Z.~Xu, M.~Yamashita, J.~Ye, G.~Zavattini, Y.~Zhang, T.~Zhu, and
  J.~P. Zopounidis.
\newblock {S}earch for {C}oherent {E}lastic {S}cattering of {S}olar
  $^{8}\mathrm{B}$ {N}eutrinos in the {XENON1T} {D}ark {M}atter {E}xperiment.
\newblock \emph{Phys. Rev. Lett.}, 126:\penalty0 091301, Mar 2021.
\newblock \doi{10.1103/PhysRevLett.126.091301}.
\newblock URL \url{https://link.aps.org/doi/10.1103/PhysRevLett.126.091301}.

\bibitem[Agnes et~al.(2023)Agnes, Albuquerque, Alexander, Alton, Ave, Back,
  Batignani, Biery, Bocci, Bonivento, Bottino, Bussino, Cadeddu, Cadoni,
  Calaprice, Caminata, Canci, Caravati, Cargioli, Cariello, Carlini,
  Cataudella, Cavalcante, Cavuoti, Chashin, Chepurnov, Cical\`o, Covone,
  D'Angelo, Davini, De~Candia, De~Cecco, De~Filippis, De~Rosa, Derbin, Devoto,
  D'Incecco, Dionisi, Dordei, Downing, D'Urso, Fiorillo, Franco, Gabriele,
  Galbiati, Ghiano, Giganti, Giovanetti, Goretti, Grilli~di Cortona, Grobov,
  Gromov, Guan, Gulino, Hackett, Herner, Hessel, Hosseini, Hubaut, Hungerford,
  Ianni, Ippolito, Keeter, Kendziora, Kimura, Kochanek, Korablev, Korga,
  Kubankin, Kuss, La~Commara, Lai, Li, Lissia, Longo, Lychagina, Machulin,
  Mapelli, Mari, Maricic, Messina, Milincic, Monroe, Morrocchi, Mougeot,
  Muratova, Musico, Nozdrina, Oleinik, Ortica, Pagani, Pallavicini, Pandola,
  Pantic, Paoloni, Pelczar, Pelliccia, Piacentini, Pocar, Poehlmann, Pordes,
  Poudel, Pralavorio, Price, Ragusa, Razeti, Razeto, Renshaw, Rescigno, Rode,
  Romani, Sablone, Samoylov, Sands, Sanfilippo, Sandford, Savarese, Schlitzer,
  Semenov, Shchagin, Sheshukov, Skorokhvatov, Smirnov, Sotnikov, Stracka,
  Suvorov, Tartaglia, Testera, Tonazzo, Unzhakov, Vishneva, Vogelaar, Wada,
  Wang, Wang, Westerdale, Wojcik, Xiao, Yang, and Zuzel]{DarkSide50:2022}
P.~Agnes, I.~F.~M. Albuquerque, T.~Alexander, A.~K. Alton, M.~Ave, H.~O. Back,
  G.~Batignani, K.~Biery, V.~Bocci, W.~M. Bonivento, B.~Bottino, S.~Bussino,
  M.~Cadeddu, M.~Cadoni, F.~Calaprice, A.~Caminata, N.~Canci, M.~Caravati,
  N.~Cargioli, M.~Cariello, M.~Carlini, V.~Cataudella, P.~Cavalcante,
  S.~Cavuoti, S.~Chashin, A.~Chepurnov, C.~Cical\`o, G.~Covone, D.~D'Angelo,
  S.~Davini, A.~De~Candia, S.~De~Cecco, G.~De~Filippis, G.~De~Rosa, A.~V.
  Derbin, A.~Devoto, M.~D'Incecco, C.~Dionisi, F.~Dordei, M.~Downing,
  D.~D'Urso, G.~Fiorillo, D.~Franco, F.~Gabriele, C.~Galbiati, C.~Ghiano,
  C.~Giganti, G.~K. Giovanetti, A.~M. Goretti, G.~Grilli~di Cortona, A.~Grobov,
  M.~Gromov, M.~Guan, M.~Gulino, B.~R. Hackett, K.~Herner, T.~Hessel,
  B.~Hosseini, F.~Hubaut, E.~V. Hungerford, An. Ianni, V.~Ippolito, K.~Keeter,
  C.~L. Kendziora, M.~Kimura, I.~Kochanek, D.~Korablev, G.~Korga, A.~Kubankin,
  M.~Kuss, M.~La~Commara, M.~Lai, X.~Li, M.~Lissia, G.~Longo, O.~Lychagina,
  I.~N. Machulin, L.~P. Mapelli, S.~M. Mari, J.~Maricic, A.~Messina,
  R.~Milincic, J.~Monroe, M.~Morrocchi, X.~Mougeot, V.~N. Muratova, P.~Musico,
  A.~O. Nozdrina, A.~Oleinik, F.~Ortica, L.~Pagani, M.~Pallavicini, L.~Pandola,
  E.~Pantic, E.~Paoloni, K.~Pelczar, N.~Pelliccia, S.~Piacentini, A.~Pocar,
  D.~M. Poehlmann, S.~Pordes, S.~S. Poudel, P.~Pralavorio, D.~D. Price,
  F.~Ragusa, M.~Razeti, A.~Razeto, A.~L. Renshaw, M.~Rescigno, J.~Rode,
  A.~Romani, D.~Sablone, O.~Samoylov, W.~Sands, S.~Sanfilippo, E.~Sandford,
  C.~Savarese, B.~Schlitzer, D.~A. Semenov, A.~Shchagin, A.~Sheshukov, M.~D.
  Skorokhvatov, O.~Smirnov, A.~Sotnikov, S.~Stracka, Y.~Suvorov, R.~Tartaglia,
  G.~Testera, A.~Tonazzo, E.~V. Unzhakov, A.~Vishneva, R.~B. Vogelaar, M.~Wada,
  H.~Wang, Y.~Wang, S.~Westerdale, M.~M. Wojcik, X.~Xiao, C.~Yang, and
  G.~Zuzel.
\newblock {S}earch for low-mass dark matter {WIMP}s with 12 ton-day exposure of
  {D}ark{S}ide-50.
\newblock \emph{Phys. Rev. D}, 107:\penalty0 063001, Mar 2023.
\newblock \doi{10.1103/PhysRevD.107.063001}.
\newblock URL \url{https://link.aps.org/doi/10.1103/PhysRevD.107.063001}.

\bibitem[Meng et~al.(2021)Meng, Wang, Tao, Abdukerim, Bo, Chen, Chen, Chen,
  Cheng, Cheng, Cui, Fan, Fang, Fu, Fu, Geng, Giboni, Gu, Guo, Han, He, He,
  Huang, Huang, Huang, Hou, Ji, Ju, Li, Li, Li, Li, Lin, Liu, Lu, Luo, Ma, Ma,
  M., Shaheed, Ning, Qi, Qian, Ren, Shang, Shen, Si, Sun, Tan, Wang, Wang,
  Wang, Wang, Wang, Wang, Wang, Wu, Wu, Xia, Xiao, Xiao, Xie, Yan, Yan, Yang,
  Yang, Yu, Yuan, Yuan, Zhang, Zhang, Zhang, Zhang, Zhao, Zheng, Zhou, Zhou,
  Zhou, and Zhou]{PandaX-4T:2022}
Y.~Meng, Z.~Wang, Y.~Tao, A.~Abdukerim, Z.~Bo, W.~Chen, X.~Chen, Y.~Chen,
  C.~Cheng, Y.~Cheng, X.~Cui, Y.~Fan, D.~Fang, C.~Fu, M.~Fu, L.~Geng,
  K.~Giboni, L.~Gu, X.~Guo, K.~Han, C.~He, J.~He, D.~Huang, Y.~Huang, Z.~Huang,
  R.~Hou, X.~Ji, Y.~Ju, C.~Li, M.~Li, S.~Li, S.~Li, Q.~Lin, J.~Liu, X.~Lu,
  L.~Luo, W.~Ma, Y.~Ma, Yajun M., N.~Shaheed, X.~Ning, N.~Qi, Z.~Qian, X.~Ren,
  C.~Shang, G.~Shen, L.~Si, W.~Sun, A.~Tan, A.~Wang, M.~Wang, Q.~Wang, S.~Wang,
  S.~Wang, W.~Wang, X.~Wang, M.~Wu, W.~Wu, J.~Xia, M.~Xiao, X.~Xiao, P.~Xie,
  B.~Yan, X.~Yan, J.~Yang, Y.~Yang, C.~Yu, J.~Yuan, Y.~Yuan, D.~Zhang,
  M.~Zhang, P.~Zhang, T.~Zhang, L.~Zhao, Q.~Zheng, J.~Zhou, N.~Zhou, X.P. Zhou,
  and Y.~Zhou.
\newblock {D}ark {M}atter {S}earch {R}esults from the {P}anda{X-4T}
  {C}ommissioning {R}un.
\newblock \emph{Phys. Rev. Lett.}, 127:\penalty0 261802, Dec 2021.
\newblock \doi{10.1103/PhysRevLett.127.261802}.
\newblock URL \url{https://link.aps.org/doi/10.1103/PhysRevLett.127.261802}.

\bibitem[Bernabei et~al.(2013)Bernabei, Belli, Cappella, Caracciolo,
  Castellano, Cerulli, Dai, d’Angelo, d’Angelo, Di~Marco, He, Incicchitti,
  Kuang, Ma, Montecchia, Prosperi, Sheng, Wang, and Ye]{DAMA/I:2013}
R.~Bernabei, P.~Belli, F.~Cappella, V.~Caracciolo, S.~Castellano, R.~Cerulli,
  C.~J. Dai, A.~d’Angelo, S.~d’Angelo, A.~Di~Marco, H.~L. He,
  A.~Incicchitti, H.~H. Kuang, X.~H. Ma, F.~Montecchia, D.~Prosperi, X.~D.
  Sheng, R.~G. Wang, and Z.~P. Ye.
\newblock {{F}inal model independent result of {DAMA/LIBRA}-phase1}.
\newblock \emph{Eur. Phys. J. C}, 73:\penalty0 2648, 2013.
\newblock \doi{10.1140/epjc/s10052-013-2648-7}.
\newblock URL \url{https://dx.doi.org/10.1140/epjc/s10052-013-2648-7}.

\bibitem[Ruppin et~al.(2014)Ruppin, Billard, Figueroa-Feliciano, and
  Strigari]{Neutrino_Floor_Xe:2014}
F.~Ruppin, J.~Billard, E.~Figueroa-Feliciano, and L.~Strigari.
\newblock {C}omplementarity of dark matter detectors in light of the neutrino
  background.
\newblock \emph{Phys. Rev. D}, 90:\penalty0 083510, Oct 2014.
\newblock \doi{10.1103/PhysRevD.90.083510}.
\newblock URL \url{https://link.aps.org/doi/10.1103/PhysRevD.90.083510}.

\end{thebibliography}

\end{document}